\PassOptionsToPackage{table,xcdraw}{xcolor}
\documentclass[sigconf]{acmart}
\AtBeginDocument{%
  }

\copyrightyear{2026}
\acmYear{2026}
\setcopyright{cc}
\setcctype{by}
\acmConference[UIST '26]{The 39th Annual ACM Symposium on User Interface Software and Technology}{November 02--05, 2026}{Detroit, MI, USA}
\acmBooktitle{The 39th Annual ACM Symposium on User Interface Software and Technology (UIST '26), November 02--05, 2026, Detroit, MI, USA}
\acmDOI{10.1145/3830398.3830481}
\acmISBN{979-8-4007-2856-3/2026/11}

\usepackage{booktabs}
\usepackage{multirow}
\usepackage{enumitem}

\newcommand{\link}[1]{%
   \texttt{\textcolor{linkblue}{#1}}%
}

\definecolor{linkblue}{HTML}{3377FF}
\newcommand{\degree}{\ensuremath{^\circ}} 

\begin{document}

\title{GazeTune: Facilitating Precise Gaze-Driven Interactions with Cascaded Touch Input}


\author{Jina Kim}
\orcid{0000-0003-3878-0818}
\affiliation{%
  \institution{KAIST}
  \city{Daejeon}
  \state{}
  \country{Republic of Korea}
}
\email{jina1190@kaist.ac.kr}

\author{Eric J. Gonzalez}
\affiliation{%
  \institution{Google}
  \city{Seattle}
  \state{Washington}
  \country{USA}
  }
  \email{ejgonz@google.com}

\author{Yang Zhang}
\affiliation{%
  \institution{University of California, Los Angeles}
  \city{Los Angeles}
  \state{California}
  \country{USA}
  }
  \email{yangzhang@ucla.edu}
  
\author{Sang Ho Yoon}
\affiliation{%
  \institution{KAIST}
  \city{Daejeon}
  \state{}
  \country{Republic of Korea}
}
\email{sangho@kaist.ac.kr}

\renewcommand{\shortauthors}{Kim et al.}

\begin{abstract}
Eye gaze has become an essential input for spatial computing, but its coarse targeting and saccadic nature limit precision and complicate continuous interactions such as dragging, especially under user motion. Gaze+pinch has also become standard in XR for its convenience, yet mid-air gestures remain imprecise, fatiguing, and socially unacceptable. These limitations underscore the need for an approach that preserves the speed of gaze while enabling stable, fine control. We present \textit{GazeTune}, a cascaded multimodal interaction technique combining gaze and touch to refine gaze-based selection and manipulation. Touch serves as a refinement channel within gaze pointing, allowing precise cursor and target control. Our work investigates how gaze-and-touch enhances dragging and mitigates \textit{Motion-Induced} instability. In a study~(N=20), we compared \textit{GazeTune} against gaze-only and gaze-pinch methods in 2D dragging. Results show that \textit{GazeTune} achieves significantly lower error with comparable execution time, validating its effectiveness and balanced trade-off between time and accuracy.
\end{abstract}



\begin{teaserfigure}
  \includegraphics[width=\linewidth]{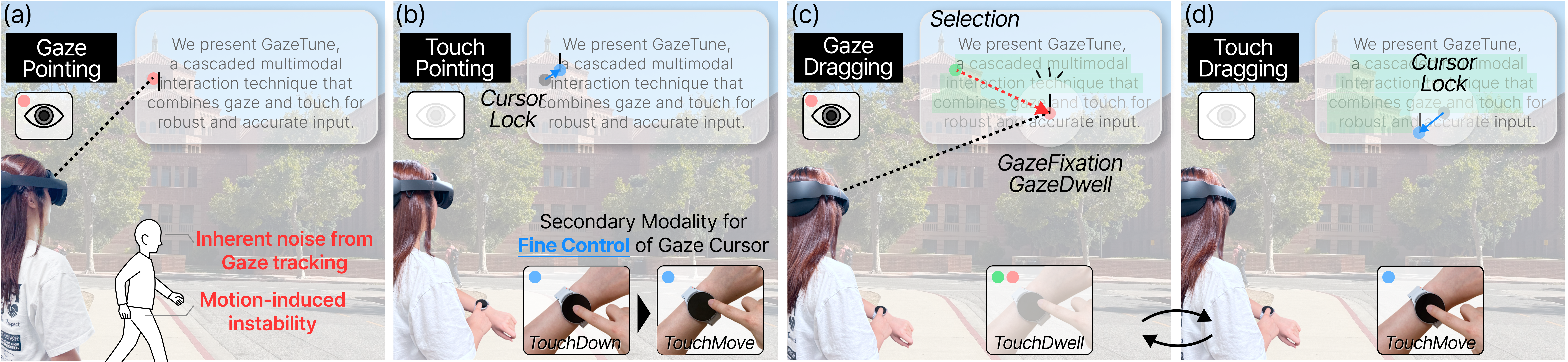}
  \caption{
  \textit{GazeTune} is a cascaded multimodal interaction combining eye gaze and touch input to overcome natural gaze imprecision and motion-induced instability. Shown here in a text selection workflow: (a)~The user coarsely positions the cursor via gaze. (b)~Touching the screen locks the cursor, allowing precise refinement via subtle finger movement. (c)~To drag an element, the user first dwells their finger to begin selection and then shifts their gaze to the coarse target destination. (d)~A new gaze fixation re-locks the cursor for a final touch refinement before finalizing placement.
  }
  \Description{Four side-by-side panels show a user wearing a head-mounted display and interacting with a floating text interface. The left pair focuses on pointing: the first panel shows coarse gaze pointing with callouts indicating gaze noise and \textit{Motion-Induced} instability, and the second shows cursor lock followed by local refinement through touch input on a smartwatch. The right pair applies the same idea to dragging: gaze supports coarse dragging across the interface, and touch dragging refines the dragged position after lock. The figure introduces the main design idea of the technique: gaze is used for fast acquisition and gross movement, and touch is added for stable fine control.}
  \label{fig:teaser}
\end{teaserfigure}


\maketitle

\section{Introduction}
Eye gaze has been considered a natural input modality for extended reality (XR), enabling fast and low-effort interaction through a simple glance.
Still, gaze input does not support continuous operations such as dragging, which demands sustained precision and stability.
While gaze is effective for coarse positioning, it lacks the stability needed for fine control due to its saccadic nature and tracking noise.
Meanwhile, gaze+pinch interactions~\cite{pfeuffer_gaze_2017} have emerged as a convenient and intuitive solution in XR.
Yet, mid-air hand gestures remain imprecise, fatiguing, and socially intrusive~\cite{hsieh2016designing,palmeira2023quantifying}.
As XR devices become increasingly portable~(e.g., AR glasses), interactions are expected to function reliably not only in stationary settings but also under locomotion. In such motion-induced conditions, it remains unclear whether gaze+pinch alone can provide stable and precise manipulation.
These limitations become particularly evident in dragging, such as adjusting sliders or specifying ranges, where continuous and reliable control is essential. 
Dragging, therefore, is not merely a stationary interaction but a fundamental operation that extends to use, where precision should be maintained in motion.
Rather than extreme movements, this motion refers to everyday urban mobility such as walking or commuting, where users are typically forced to stop due to gait-induced noise when performing such precision tasks.

Several multimodal techniques combine gaze with complementary inputs~(e.g., head~\cite{sidenmark_bimodalgaze_2020, kyto_pinpointing_2018}, hand~\cite{chatterjee_gazegesture_2015, wagner_eye-hand_2024}, controller~\cite{chen_gazeraycursor_2023}, touch~\cite{stellmach_look_2012, turner_gazerst_2015, cai_gazeswipe_2025}) to compensate for its inherent imprecision.
While effective in stationary settings, these approaches often struggle to support fine-grained corrections during physical movement, as they tend to rely on discrete steps rather than fluid, small-amplitude adjustments.
A recent study~\cite{wagner_eye-hand_2024} demonstrated the potential of gaze-assisted dragging with fluid modality transitions, but it did not explicitly address interaction while the user is in motion.
Furthermore, as XR moves toward everyday use, users will need precise manipulation techniques that remain practical in public spaces where large, mid-air gestures are undesirable.
This highlights the value of leveraging companion devices, like smartwatches, to enable subtle corrective input without adding physical strain or social intrusiveness.

These considerations motivate a gaze-driven design that preserves rapid coarse targeting while enabling fine, continuous manipulation with minimal physical effort. 
To achieve this, we leverage a smartwatch as a low-overhead, always-accessible complementary channel for subtle touch corrections. 
Our goal is to combine rapid gaze acquisition with stable touch refinement within a single continuous sequence, reducing physical effort and ensuring reliable control.
To this end, we introduce \textit{GazeTune}, a cascaded interaction technique that pairs coarse gaze targeting with precise touch refinement using a commercial smartwatch. The interaction unfolds in four phases: a)~setting the coarse start point with gaze, b)~locking and refining it via touch, c)~specifying a rough endpoint with gaze, and d)~refining the final endpoint with touch. Through this flow, gaze offers rapid spatial anchoring while touch enables localized fine-tuning, enabling a natural and reliable dragging experience that neither modality could achieve alone.

To validate our approach, we conducted a user study comparing \textit{GazeTune} against two baselines: \textit{GazePinch} (which represents the state-of-the-art interaction widely used in XR) and \textit{GazeTap} (which captures the inherent limitations of gaze-only dragging).
We implemented a Fitts’ law-inspired, consecutive selection-and-dragging task across two contexts (\textit{Stationary} vs. \textit{Motion-Induced}) to systematically measure efficiency, accuracy, and workload with a specific focus on user mobility.
We found that under the \textit{Motion-Induced} condition, \textit{GazeTune} achieved a significantly lower error rate (4.24\%) compared to \textit{GazePinch} (37.12\%) and \textit{GazeTap} (9.70\%), while maintaining comparable task completion times. Furthermore, \textit{GazeTune} demonstrated greater stability during manipulation; it yielded faster Drag \& Drop times than \textit{GazePinch} (2.59~s vs. 4.09~s) and resulted in a lower dropping error angle (1.44°) than both \textit{GazeTap} (1.89°) and \textit{GazePinch} (2.91°). Overall, our findings demonstrate that \textit{GazeTune} effectively minimizes errors and user workload while preserving the speed advantages of gaze input. This underscores its resilience during user locomotion and validates our seamless modality-switching approach. Overall, these results reveal how users adapt to gaze-and-touch integration, providing insights for future designs that balance rapid targeting with precise correction.

In summary, this work makes three contributions:
\begin{itemize}
\item A novel cascaded gaze-and-touch interaction mechanism for precise dragging in XR.
\item An empirical evaluation comparing \textit{GazeTune} against existing gaze-driven baselines, validating its ability to enhance interaction precision and stability.
\item Design implications for integrating touch input into mobile AR contexts to extend the capabilities of current gaze-based interactions.
\end{itemize}


\section{Related Work}
\subsection{Gaze-Driven Interaction Techniques}
In this subsection, we highlight why refinement is necessary for gaze-driven manipulation. Eye movement-based interaction~\cite{jacob_what_1990, tanriverdi_interacting_2000} has become a widely adopted input modality across both 2D devices and immersive XR environments. Because gaze naturally reflects a user's attention and intent to manipulate objects~\cite{wagner_eye-hand_2024}, it is often perceived as faster and less physically demanding than other inputs. To further enhance its capabilities, researchers have explored multimodal interactions that combine gaze with additional modalities to improve expressiveness and input accuracy. For example, gaze has been paired with hand gestures~\cite{pfeuffer_gaze_2017, yu_gaze-supported_2021, shi_exploring_2023, jeong_gazehand_2023}, as well as touch input for tasks like mode switching~\cite{pfeuffer_gaze-touch_2014, pfeuffer_gaze_2016, gazebutton} and text entry~\cite{sindhwani_retype_2019, kumar_tagswipe_2020}. While these complementary strategies effectively mitigate gaze ambiguity in contexts such as cross-device positioning~\cite{turner_cross-device_2014}, occluded-object selection~\cite{xu_eyeexpand}, depth-aware targeting~\cite{zhang_focusflow_2023, chen_gazeraycursor_2023}, and voice interaction~\cite{zhao_eyesaycorrect_2022}, the majority of these evaluations focus almost exclusively on precise selection.

Consequently, the potential of multimodal gaze systems for continuous operations~(e.g., selection and manipulation~\cite{velloso_empirical_2015}) - which require sustained adjustments and stable post-selection dragging during physical movement - remains underexplored. We address this gap with \textit{GazeTune} by integrating touch into a cascaded refinement workflow. By pairing natural gaze pointing with subtle touch corrections, our approach enables stable and effective dragging even while the user is in motion.

\subsection{Multimodal Refinement for Gaze Input}
Even with advanced tracking and calibration~\cite{cai_gazeswipe_2025}, gaze alone is often insufficient for precise target selection and remains inherently error-prone~\cite{stellmach_look_2012}. Beyond technical inaccuracies, gaze-based systems suffer from the ``Midas touch'' problem, which leads to ambiguous or unintended selections. To address this, prior work has explored multimodal techniques combining gaze with complementary inputs~(e.g., head, hand, or touch) to improve accuracy and reliability. Among these, pairing gaze with touch has proven highly effective~\cite{turner_gazerst_2015, stellmach_look_2012, stellmach_still_2013}, providing users with the explicit control needed to correct errors on the fly. While other complementary modalities offer advantages, they face limitations for precise refinement. Head-based input~\cite{sidenmark_bimodalgaze_2020, kyto_pinpointing_2018, lookandlean} is intuitive, yet turning the head can be inconvenient and visually distracting. Mid-air gestures~\cite{chatterjee_gazegesture_2015, deng_understanding_2017, kyto_pinpointing_2018, SuEnhanced2025, wagner_eye-hand_2024, H2H} are versatile but become unreliable for fine refinement during physical movement. Devices like the mouse (via MAGIC~\cite{zhai_manual_1999, fares_can_2013}) or controllers~\cite{chen_gazeraycursor_2023} support high precision but lack portability. Finally, approaches relying on voluntary eye movements~\cite{eyelid3d_2022} or error-aware systems~\cite{Weightedpointer, guarnera_automated_2025} help mitigate inaccuracies but cannot fully substitute for deliberate, user-controlled correction.

Given these limitations, we believe wearable touch offers a practical input channel for precise, selective refinement in mobile contexts. Devices such as smartwatches offer readily available, unobtrusive touch surfaces that integrate into everyday activities without requiring users to hold external hardware or perform large, fatiguing body movements. As such, wearable touch serves as an ideal complement to gaze, providing the stability necessary for fine-grained control on the go.

\subsection{Impact of Mobility on User Interactions}
Prior work has addressed position- and movement-independent gaze interaction on large public displays~\cite{khamis_eyescout_2017}, characterized gaze accuracy and precision ranges on 2D screens~\cite{ref:Toward_everyday_gaze}, and explored continuous eye-movement control via smooth-pursuit adjustment tasks~\cite{niu_smooth_2023}. 
While gaze has long been considered a powerful input modality, its serious consideration in mobile contexts is relatively recent~\cite{gazeonthego}.
Studies have also examined how user movement and task context shape the effectiveness of input.
They revealed that locomotion and body motion impair input performance on HMDs across hand-, head-, and eye-based techniques, leading to higher error and latency while increasing attention and motor demands~\cite{li_evaluating_2024, li_estimating_2025, shin_using_2025}.
Furthermore, the necessity of gaze compensation in mobile is highlighted by work showing that the choice of spatial reference critically affects tracking precision~\cite{gazeonthego}.
These findings underscore the importance of considering situational impairments but remain focused on discrete acquisition tasks rather than continuous manipulation. 

Dragging, however, is prone to motion-induced physical instability, leading to jitter and drift along the path that requires maintaining control across the entire sequence~(start point, path, and end point). 
Each stage can be affected by movement-related disturbances, yet dragging has received comparatively little attention despite its practical relevance in XR.
To address this gap, we focus on stabilizing gaze-driven dragging under motion-induced conditions and evaluate how a gaze~\&~touch technique supports stable acquisition and intentional real-time refinement.

\begin{figure}[t]
    \centering
    \includegraphics[width=0.99\columnwidth]{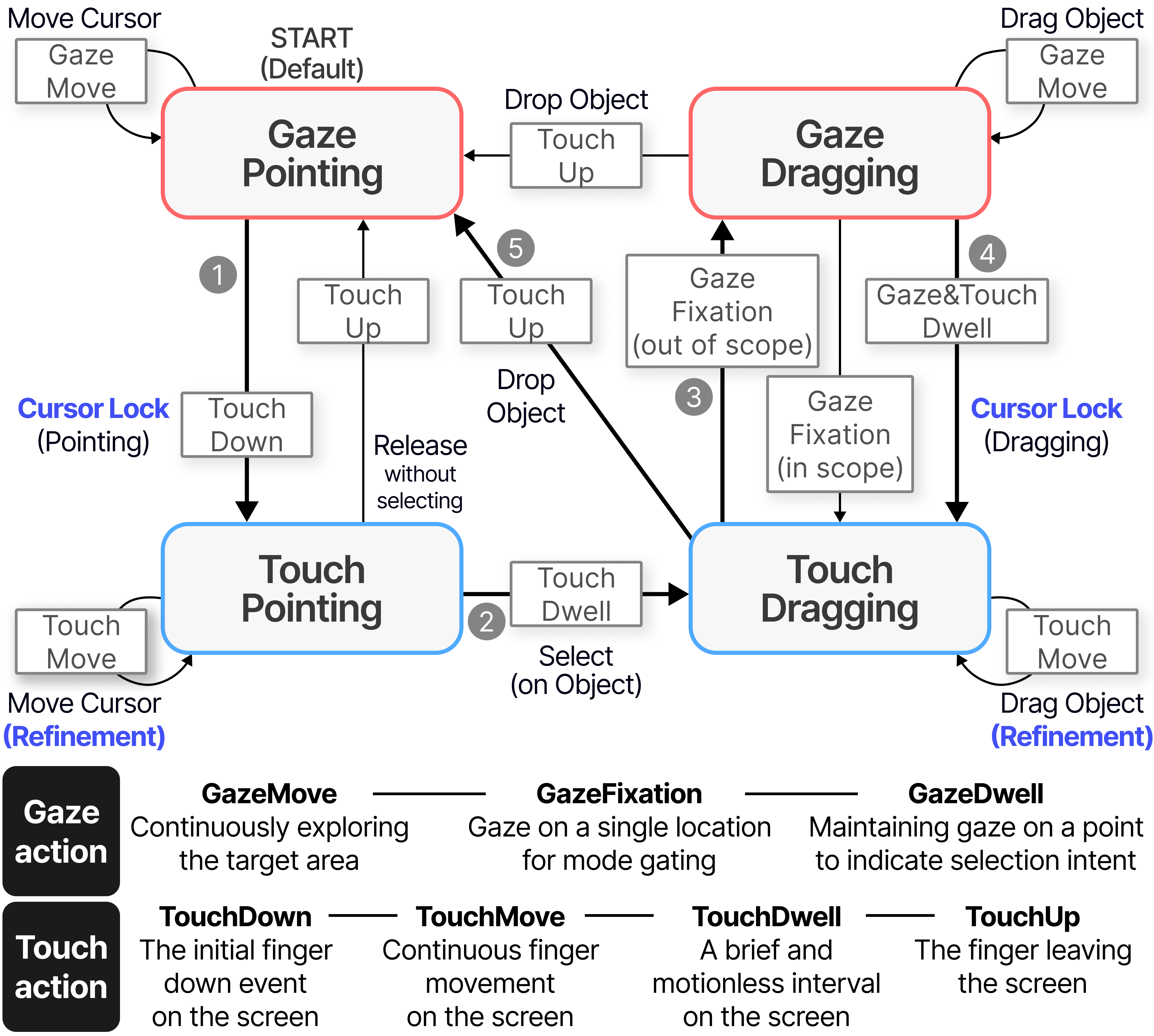}
    \Description{Interaction state machine for GazeTune. The diagram contains four main states arranged in a two-by-two layout: Gaze Pointing and Touch Pointing on the left, and Gaze Dragging and Touch Dragging on the right. Interaction begins in Gaze Pointing, where gaze controls coarse cursor movement. Touch-down transitions to Touch Pointing, which represents cursor lock and local refinement through touch movement. Touch dwell on an object initiates dragging and moves the interaction to the dragging side. On the right, Gaze Dragging represents coarse movement of the selected object, and Touch Dragging represents fine refinement after lock. The system switches between these two depending on gaze conditions: fixation within scope or coordinated dwell enters touch-based drag refinement, and fixation outside scope returns control to gaze-based dragging. The diagram summarizes the event-driven modality switches used throughout the interaction.}
    \caption{The interaction diagram of GazeTune based on the gaze and touch actions. All processes can be performed within a single touch sequence to support two main phases of refinement~(Pointing and Dragging). The numbered arrows (1-5) illustrate the primary interaction flow. Any TouchUp event ends the interaction cycle, returning the system to Gaze Pointing.}
    \label{fig:interaction_state}
\end{figure}
\setlength{\textfloatsep}{6pt} 

\section{GazeTune System Design}
\subsection{Design Objectives: Gaze-Driven Dragging with Minimal Touch Input}
In this section, we outline the design goals of \textit{GazeTune}.
Our motivation is to enable reliable gaze-driven dragging by combining touch with the rapid targeting advantages of gaze, particularly in  scenarios. 
However, when users are in motion, they cannot maintain a stable device posture or devote sustained visual attention~\cite{brewster_multimodal_2003}.
Under these conditions, our design focuses on enabling low-attention interaction that operates effectively with minimal movement on compact touch surfaces while complementing gaze interaction across a wide XR space.
Building on this approach, we pair gaze with subtle touch refinement to support low-effort interaction and natural modality transitions, leading to the following design goals~(G):
\begin{itemize}
    \item[\textbf{[G1]}] \textbf{Leverage Input Familiarity}: To reduce the learning effort, the interaction should align with familiar cursor manipulation metaphors~(e.g., trackpad- or mouse-based selection)
    \item[\textbf{[G2]}] \textbf{Enable Natural Modality Transitions}: Switching between gaze and touch should be seamless to achieve high precision with low cognitive load and minimal physical effort.
    \item[\textbf{[G3]}] \textbf{Extend Drag Reach from Compact Touch Surfaces}: The technique should remain usable on small touch surfaces typical of wearables, enabling distant selection while minimizing physical clutching.
\end{itemize}


\begin{figure*}[t!]
    \centering
    \includegraphics[width=0.9\linewidth]{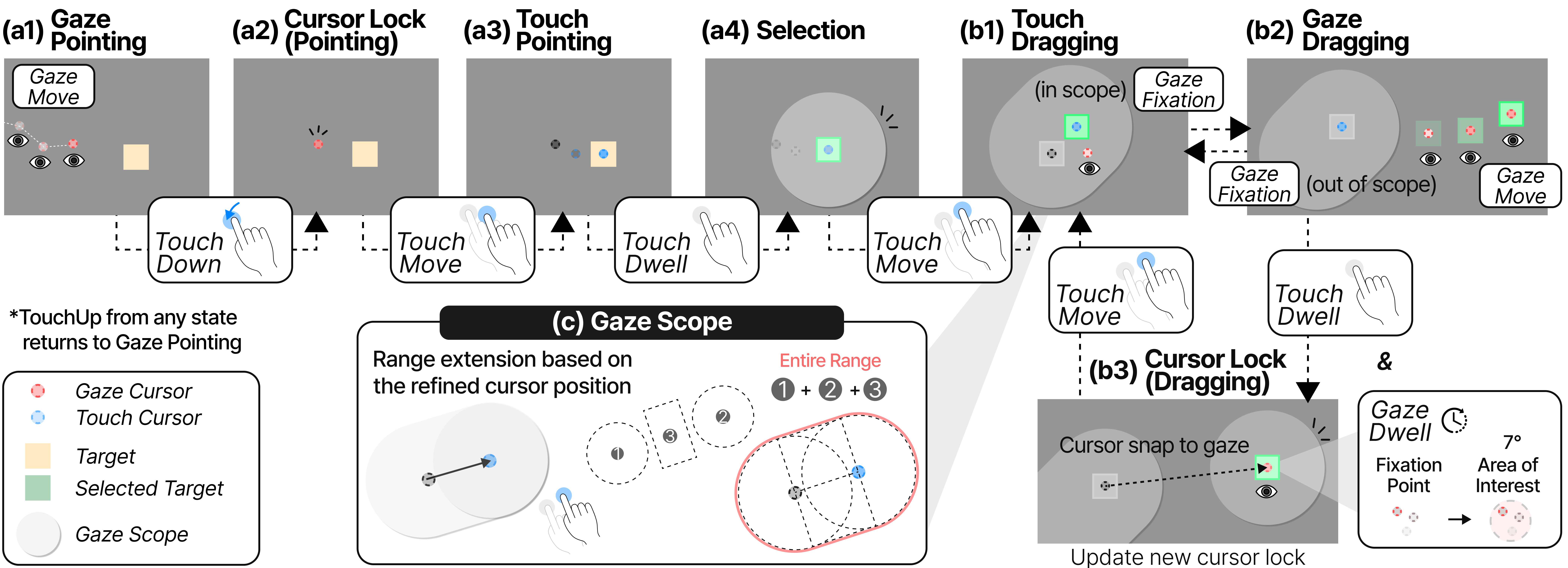}
    \Description{Interaction mechanism for pointing, dragging, and adaptive scope. The figure is divided into three parts. The first part shows a pointing-and-selection sequence in which the user explores with gaze, locks the cursor with touch-down, refines the position with touch movement, and confirms the target with touch dwell. The second part shows dragging, where the user performs local touch refinement while gaze remains within a valid scope, switches back to gaze-based coarse retargeting when the intended destination lies outside that scope, and then establishes a new lock position through later fixation. The third part explains that the valid gaze scope is not fixed: it expands according to the refined cursor position, covering the original lock point, the refined point, and the area between them. The figure explains both the action flow and the control logic that allows gaze and touch to alternate smoothly.}
    \caption{(a1)~Gaze pointing to explores the approximate area. (a2)~A TouchDown creates a Cursor Lock to anchor the cursor. (a3)~Touch pointing to refine the cursor position relative to the lock. (a4)~A TouchDwell confirms the selection once the cursor reaches the target. (b1)~Touch dragging occurs when the gaze remains within the predefined scope. (b2-3)~Gaze dragging occurs when the gaze moves beyond the scope and satisfies fixation-based rules, allowing the cursor to snap to a new location for continued touch dragging. The system can repeat iteratively between two stages based on gaze position. (c)~The scope is dynamically updated based on the refined cursor position. }
    \label{fig:mechanism}
\end{figure*}

\subsection{Overview \& Workflow}
We designed a cascaded interaction that combines coarse gaze targeting with fine-grained touch control in a single continuous sequence. The \textit{GazeTune} is realized through three gaze actions and four touch actions~(Figure~\ref{fig:interaction_state}) for state transitions. Regarding \textit{TouchDwell}, we found that 300 ms was a feasible dwell time, as the touch already indicates intentional use. These events support two linked interaction sequences: Point \& Select for target acquisition and Drag \& Drop for object manipulation. 

\begin{enumerate}[leftmargin=*]
    \item \textbf{Gaze Pointing}: As illustrated in Figure~\ref{fig:teaser}, the user first explores the approximate area with gaze for coarse target acquisition~\cite{kumar_eyepoint_2007, turner_eye_2013-1, turner_eye_2013}.

    \item \textbf{Touch Pointing}: A \textit{TouchDown} anchors the cursor at the current gaze position. We refer to this anchor as a \textit{Cursor Lock}. The user can \textit{TouchMove} to refine the cursor relative to this lock, enabling subsequent selection and dragging~(G1). 
    
    \item \textbf{Select}: When the cursor reaches the target, a \textit{TouchDwell} confirms the selection for subsequent manipulation. For Point \& Select, the interaction ends with \textit{TouchUp}.
\end{enumerate}

Our system maintains a \textit{Gaze Scope} around the current lock to regulate transitions. The scope functions as a gating region that determines whether control remains in local touch refinement or transitions to gaze targeting~(G2).

\begin{enumerate}[leftmargin=*]
\setcounter{enumi}{3}
    \item \textbf{Touch Dragging}: After selection, the user can \textit{TouchMove} to drag the object while gaze remains within the Gaze Scope. 
    
    \item \textbf{Gaze Dragging}: When a larger repositioning is needed, the user shifts gaze outside the Gaze Scope.
    While maintaining a \textit{TouchDwell} and \textit{GazeDwell}, a fixation establishes a cursor lock~(dragging) for further refinement where the cursor \textit{snaps} to the current gaze, similar to prior work~\cite{turner_gazerst_2015, yu_gaze-supported_2021}.
    If the fixation returns within the Gaze Scope, the system returns to \textit{Touch Dragging}. 

    \item \textbf{Drop}: The interaction is completed by \textit{TouchUp}, dropping the object at the final location.
\end{enumerate}

By delegating large-scale tasks to gaze and reserving touch for short and anchored refinements, it supports precise local refinement and rapid repositioning within a single sequence that could minimize clutching~(G3).

\begin{figure*}[t]
    \centering
    \includegraphics[width=0.96\linewidth]{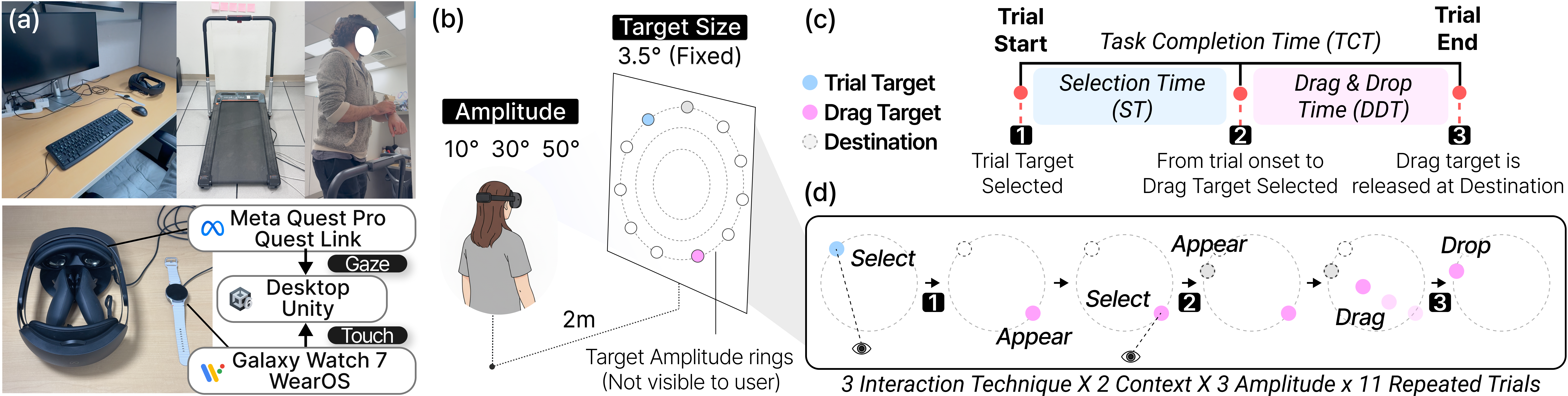}
    \caption{(a)~Study setup for the \textit{Stationary} and \textit{Motion-Induced} conditions. Participants walk on the treadmill while wearing an HMD~(Meta Quest Pro, 1800 × 1920 pixels per eye, 90 Hz refresh rate) and controlling a smartwatch touchscreen~(Galaxy Watch 7, SM-L310, 44~mm). (b)~Overall study setup. The trial target appears in a circular layout, and the drag target appears if the user selects it. (c)~Task time subcomponents and procedure of each trial. (d)~Each trial begins with selecting the trial target. From trial onset to drag target selection is measured as Selection Time~(ST). From drag target selection to dropping it at the destination is the Drag~\&~Drop Time~(DDT). The total duration is Task Completion Time~(TCT). Trials repeat across techniques, contexts, and amplitudes.}
    \Description{User study apparatus, target layout, and task sequence. The figure combines the experiment hardware and trial design into four panels. One panel shows the apparatus, including a head-mounted display, smartwatch touch input, desktop application, and both stationary and treadmill conditions. Another shows the target layout at three amplitudes, labeled Near, Mid, and Far, with circular target positions placed on angular rings and a fixed target size. A third panel illustrates the drag-and-drop trial sequence from initial target appearance through selection, dragging, and release at the destination. The last panel summarizes the repeated-measures design across three techniques, two motion contexts, three amplitudes, and repeated trials. The figure provides a compact view of both the physical setup and the structure of a single experimental trial.}
    \label{fig:userstudy}
\end{figure*}

 \subsection{Cursor Lock for Anchored Refinement}
 We adopt a \textit{Cursor Lock} mechanism in which gaze provides a spatial anchor while touch enables local cursor refinement.
 When a Cursor Lock is activated, the gaze cursor is immediately frozen, establishing a reference point for subsequent touch-driven movements.
This design supports unobtrusive interaction in scenarios by allowing users to refine input quickly without requiring excessive reliance on prolonged visual attention.
We define two types of locks, each relevant for a different interaction phase:

(1)~\textbf{Pointing Cursor Lock (Touch)}. An initial lock is activated during coarse gaze pointing (Figure~\ref{fig:mechanism}a1) at first touch (Figure~\ref{fig:mechanism}a2). This locks the cursor and supports local refinement of the pointing cursor before selection (Figure~\ref{fig:mechanism}a3).

(2)~\textbf{Dragging Cursor Lock (Fixation)}. Activated during gaze dragging (Figure~\ref{fig:mechanism}b2). When gaze remains within a 7° radius over a 400~ms window, the system registers stable fixation.
And the refinement anchor is re-established at the new fixation point (Figure~\ref{fig:mechanism}b3).

\subsection{Gaze Scope for Mode Gating}
To ensure stable refinement and avoid unintended cursor jumps from natural saccadic eye movement, we configure a \textit{Gaze Scope}~(Figure~\ref{fig:mechanism}c), a region surrounding the refinement anchor position.
This region determines whether the user intends to perform fine-grained refinement or reposition the cursor over a long distance. 

The Gaze Scope region dynamically extends based on the refined cursor's position relative to the initial Dragging Cursor Lock anchor. It forms a flexible pill-shaped region (i.e., discorectangle) composed of two $10.1\degree$ circles centered on the initial anchor position, the current refinement cursor position, and a rectangular corridor connecting them~(Sec~\ref{implementation}).
The Gaze Scope gates the system between two distinct modes of operation during dragging interaction: (1)~\textit{Touch Dragging} and (2)~\textit{Gaze Dragging}.
While the gaze remains inside the Gaze Scope, the system is in a touch dragging mode~(Figure~\ref{fig:mechanism}b1).
Gaze is ignored for positional updates, allowing for stable manipulation.
When the gaze intentionally moves and remains outside the scope, the system transitions to coarse gaze-based dragging~(Figure~\ref{fig:mechanism}b2). 
If the user fixates on a new point for over 400 ms, a Cursor Lock~(Dragging) ``snaps'' to the current gaze position, establishing a new anchor for refinement.

This gating mechanism ensures a predictable transition between precise touch-driven manipulation and rapid gaze-assisted aiming. This reduces touchscreen clutching and supports a more seamless and efficient interaction flow.

\section{User Study}
The objective of our study is to evaluate the performance and explore the behavior of \textit{GazeTune} in comparison to two representative gaze-based techniques.
\textit{GazeTap}~(Baseline 1) represents the baseline of direct gaze pointing, and \textit{GazePinch}~(Baseline 2) is a widely adopted mid-air combination of gaze+pinch, regarded as a commonly used commercial method but prone to fatigue and jitter. 
We aim to assess both performance and trade-offs among techniques. 
Specifically, we address following research questions:

\begin{itemize}[leftmargin=*, itemsep=2pt, topsep=2pt, parsep=0pt, partopsep=0pt]
    \item \textbf{RQ1:} How does \textit{GazeTune} compare to  \textit{\textit{GazeTap}} and  \textit{\textit{GazePinch}} in overall performance, and what advantages does it provide in the selection and dragging tasks?
    \item \textbf{RQ2:} How does the integration of touch input influence performance under different mobility conditions?
    \item \textbf{RQ3:} How do users adapt their behavior when using GazeTune, and how do they perceive its workload and usability?
\end{itemize}

\begin{table}[t]
\caption{Comparison table for input of each technique. The primary modality for pointing is eye gaze in common.}
\resizebox{\columnwidth}{!}{
\begin{tabular}{c|c|c|c}
                      & \textbf{GazeTap} & \textbf{GazePinch} & \textbf{GazeTune~(Ours)} \\ \hline
\textbf{Select}       & TouchDown              & Pinch             & TouchDown               \\ \hline
\textbf{Drag Trigger} & TouchDwell      & Pinch-in          & TouchDwell + GazeDwell       \\ \hline
\textbf{Drag}         & Eye gaze         & Arm Movement      & Eye gaze + TouchMove  \\ \hline
\textbf{Drop}         & TouchUp             & Pinch-out         & TouchUp            
\end{tabular}
}

\label{tab:eachmethods}
\end{table}
\subsection{Study Design Rationale}
\paragraph{Smartwatch Use}
In line with controller-free XR usage, a smartwatch offers users an instantly accessible body-worn touch surface without retrieving or holding a device for unobtrusive input.
Since our technique relies on minimal, low-effort touch to refine gaze-based control, a small anchored surface is more suitable than handheld devices that require explicit grip or posture adjustments. 
Prior work shows that users naturally stabilize their interacting finger by resting it on the watch, enabling controlled and precise contact gestures~\cite{mobility_smartwatch}. 
Based on these considerations, we adopted a commercial smartwatch as a practical and reliable source of precise minimal touch input.
Despite temporarily engaging both hands, smartwatches offer social and ergonomic advantages, providing discreet interactions and avoiding the fatigue and awkwardness of mid-air gestures. Furthermore, unlike smartphones that require constant holding and retrieval, smartwatches leave hands free.

\vspace{-0.25\baselineskip}
\paragraph{GazeTap~(Touch-based Baseline)}
\textit{GazeTap} relies on gaze for pointing and dragging, and selection is confirmed by a tap. 
The user maintains finger contact while the drag position is updated by gaze, and the operation is completed by lifting the finger as illustrated in Table~\ref{tab:eachmethods}. 
This baseline helps isolate the inherent limitations of gaze-only dragging and motivates the need for touch-based refinement by showing where refinement can contribute to improved control.
\vspace{-0.25\baselineskip}
\paragraph{\textit{GazePinch} (Multimodal-based Baseline)}
\textit{GazePinch} combines gaze pointing with a pinch gesture to trigger and maintain dragging. 
Once a pinch is performed, the drag position is updated by hand movement until the pinch is released.
Although our technique adds a touch surface, \textit{GazePinch} remains an essential baseline because it is the dominant multimodal interaction on commercial HMDs and the closest deployable alternative for a continuous baseline~\cite{wagner_eye-hand_2024}.
It clarifies the benefits of combining gaze with a touch surface and the refinement workflow, providing a practical point of comparison with existing gaze-based indirect arm dragging.

\vspace{-0.25\baselineskip}
\paragraph{Treadmill Use (Motion-Induced Condition)}
We also examine how user mobility affects gaze–touch dragging. 
To validate our method in a physical mobility setting, we included a treadmill \textit{Motion-Induced} condition~\cite{headar, li_evaluating_2024, shin_using_2025}, where body motion can introduce additional gaze offset and stability challenges. 
We considered two conditions: a \textit{Stationary} condition typical of controlled user studies and a \textit{Motion-Induced} condition where the user walks at a steady pace (1.2~m/s), ensuring a consistent level of body motion across techniques to avoid speed adaptation.
We adopted a path-reference~\cite{gazeonthego}, which remained at a fixed distance and height to the user's path of locomotion. 
We evaluate whether gaze–touch integration can maintain interaction accuracy and speed even under mobility, where gaze input is more susceptible to error.

\begin{figure*}[t!]
    \centering
    \includegraphics[width=0.9\textwidth]{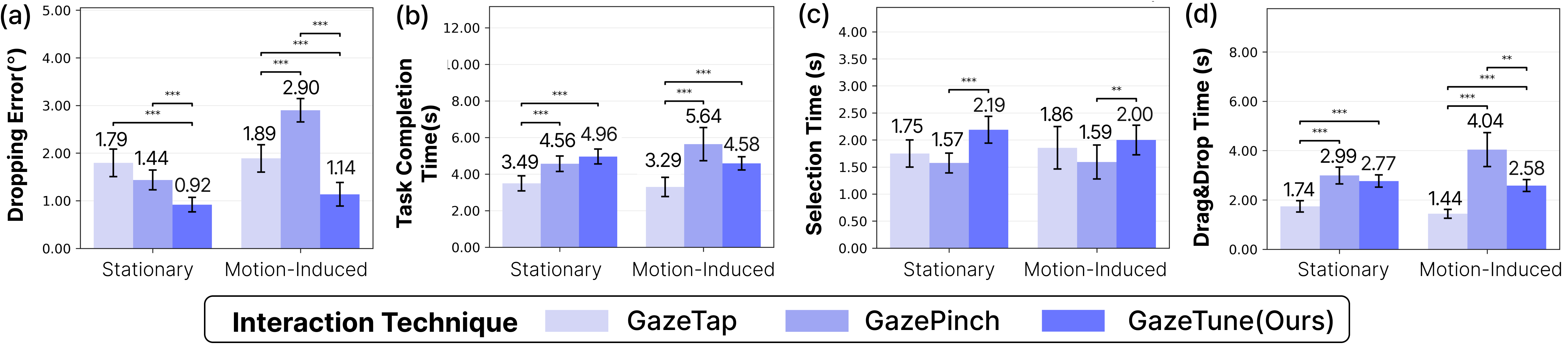}
    \caption{Results across context for each technique with the main effect of technique.}
    \Description{Technique-level performance comparison for error and time. Four grouped bar charts compare GazeTap, GazePinch, and GazeTune in \textit{Stationary} and \textit{Motion-Induced} contexts for dropping error, task completion time, selection time, and drag-and-drop time. The dropping-error chart shows the clearest separation: GazeTune has the lowest error overall, and its advantage becomes especially visible when motion is introduced, whereas GazePinch degrades sharply. The three time charts show smaller and more mixed differences, indicating that the strongest benefit of the proposed technique lies in robustness and placement accuracy.}
    \label{fig:ratings}
\end{figure*}

\subsection{Apparatus and Task}
Our user study followed ISO~9241-9~\cite{ISO9241-9} for a Fitts’-style evaluation of selection and dragging.
Targets were placed on a circular ring at a viewing distance \(2\,\mathrm{m}\) from the participant and eye height, divided into 11 equal sectors (\(\approx 32.7^\circ\)). 
A sector served as the initial index in each block, and subsequent trials are a sector on the opposing side of the previous sector.
Target amplitude was manipulated by visual angle with three levels near \((10^\circ)\), mid \((30^\circ)\), and far \((50^\circ)\), while target size was fixed at \(3.5^\circ\). 
Each trial followed a three-step sequence. (1)~select the trial target to start, which reveals a drag target, (2)~select the drag target to reveal the destination target, and (3)~drag the drag target to the destination and release it to complete. 
Participants completed 18 blocks in total, consisting of 3~interaction techniques~(\textit{GazeTap}/\textit{\textit{GazePinch}}/\textit{GazeTune})~$\times$~2~contexts~(\textit{Stationary}/\textit{Motion-Induced})~$\times$3~amplitudes~(\textit{Near/Mid/Far}). 

We implemented all techniques in Unity 6 using Meta’s built-in Interaction SDK\footnote{Meta \href{https://developers.meta.com/horizon/reference/interaction/v78/}{\link{Interaction SDK}}} for gaze tracking and pinch detection~(Figure~\ref{fig:userstudy}(a)).
Gaze rays were cast onto the task plane and smoothed with a 1€ filter to get a stable gaze~\cite{chen_gazeraycursor_2023}.
Touch events on Galaxy Watch 7\footnote{Samsung \href{https://www.samsung.com/sec/watches/galaxy-watch/galaxy-watch7/}{\link{Galaxy Watch 7}}} are delivered by \texttt{TouchDown}/\texttt{TouchMove}/\texttt{TouchUp}~provided by XDTK~\cite{GonzalezXDTK2024}.
We provided a small semi-transparent gaze cursor and displayed a gaze scope only during refinement.
We provided a short haptic pulse for the interaction stage (i.e., when a dragging lock was created) to provide an immediate response to user action using the built-in actuator in watch-enabled conditions~({GazeTap} and {GazeTune})~\cite{haptic_eyestyping, sonohaptic}.
We did not include haptic in the \textit{GazePinch} as the pinch action itself provides cutaneous cues.

\subsection{Implementation}
\label{implementation}
\paragraph{GazeTap}
Gaze performs both acquisition and continuous dragging, while touch gates selection and commitment. On \textit{TouchDown}, gaze selection occurs with a short dwell~(300~ms). While the touch is held, the selected target continuously follows the gaze without anchoring. On \textit{TouchUp}, the dragging ends, and then the stage is cleared.
\begin{equation}\label{eq:baseline}
\operatorname{pos}_{\text{target}}(t)=\operatorname{pos}_{\text{gaze}}(t)
\end{equation}

\paragraph{GazePinch}
We used \texttt{IndexPinchSelector} from Meta SDK, which provides \texttt{WhenSelected}/\texttt{WhenUnselected} events to detect pinch-in/out. On \textit{Pinch-in}, the system checks if the gaze lies within the target and, if so, enters the dragging stage and records an initial world-space offset between the target and \texttt{OculusCursor} serves as the gaze-anchored cursor proxy. During dragging, the target is moved by \texttt{OculusCursor} by controlled by arm movement while preserving the constant offset (Equation~\ref{eq:offset}). On \emph{Pinch-out}, the system exits the dragging stage and finalizes the manipulation. This follows the familiar "gaze to aim, pinch to hold" workflow while providing an indirect drag driven by gaze and hand.
\begin{equation}\label{eq:offset}
\begin{aligned}
\operatorname{pos}_{\text{target}}(t) &= \operatorname{pos}_{\text{cursor}}(t)+\Delta_{0},\\
\Delta_{0} &= \operatorname{pos}_{\text{target}}(t_{0})-\operatorname{pos}_{\text{cursor}}(t_{0}).
\end{aligned}
\end{equation}

\paragraph{GazeTune}
On \textit{TouchDown}, the system locks the current gaze position and anchors the cursor. 
During \textit{TouchMove}, selection is confirmed when the refined cursor enters the target and satisfies the dwell execution time~(same as GazeTap). 
For touch dragging, on-screen \textit{TouchMove} deltas directly drive refinement. 
For gaze dragging, the target snaps toward the current gaze if the gaze leaves the scope while \textit{TouchDwell} remains active.
The pill-shaped scope consists of two fixed-radius circles centered at the initial lock and refined cursor, plus the connecting corridor between them.
Accordingly, gaze gating is determined by three distances: the distance from the current gaze to the initial anchor, the distance to the current refined cursor, and the perpendicular distance to the connecting segment.

A fixation-based Cursor Lock is performed during a brief \textit{TouchDwell} when three requirements are met: (1) a valid fixation, (2)~the gaze is out-of-scope, and (3)~low touch speed to avoid accidental recentering. 
Once a cursor lock is established, refinement resumes from the updated anchor position.
On \textit{TouchUp}, the system ends the refinement and clears the lock.
\begin{align}
\text{Touch Dragging: }\operatorname{pos}_{\text{target}}(t) &= \operatorname{pos}_{\text{refined}}(t), \label{eq:scope_in}\\
\text{Gaze Dragging: }\operatorname{pos}_{\text{target}}(t) &= \operatorname{pos}_{\text{gaze}}(t). \label{eq:scope_out}
\end{align}

Touch input was mapped to a non-linear control--display gain, where $\Delta \mathbf{p}$ denotes the touch delta in pixel units. We computed the touch velocity as $\mathbf{v}=\Delta \mathbf{p}/\Delta t$ with speed $s=\|\mathbf{v}\|$, 
\[
g(s)=g_{\min}+\left(g_{\max}-g_{\min}\right)\left(1-e^{-s/v_{\max}}\right),
\]
where $g_{\min}=0.001$, $g_{\max}=0.003$, and $v_{\max}=250$.
The delta movement is based on smartwatch pixel level, which is scaled by CD gain and mapped to the corresponding visual angles at a 2~m viewing distance.
The scope was not a hard boundary on cursor movement but a gating region for switching control modes, remaining controllable through relative gain anchored to the initial cursor lock.

\subsection{Participants and Procedure}
We carried out an IRB-approved in-person study with 20 participants (8 females, mean age of 24), and 18 participants were right-handed, and 2 participants were left-handed. We first performed calibration for eye gaze and had a practice session to get used to each interaction technique. 
The total number of conditions was 18~(3 interaction techniques~$\times$~2 conditions~$\times$~3 target amplitudes). 
We counterbalanced the order of the task blocks for 3 different orders of interaction techniques across participants. 
Within each block, the context order was blocked (\textit{Stationary} and \textit{Motion-Induced}) to reduce early variability due to motion-induced jitter, and 3 different target amplitude levels were randomized. 
We collected a total of 3,960 data points (20 users $\times$ 3 interaction techniques $\times$ 2 contexts $\times$ 3 target amplitudes $\times$ 11 trials).

\subsection{Measures}
Our analysis considers the following metrics with task time decomposed into subcomponents as shown in Figure~\ref{fig:userstudy}.
\begin{itemize}[leftmargin=*, itemsep=2pt, topsep=2pt, parsep=0pt, partopsep=0pt]
    \item Dropping Error~(DE): Angle difference between the ray from the head to the target's final position and the ray from the head to the intended target position at the end of the trial.
    \item Task Completion Time~(TCT): Time from target appearance to dropping.
    \item Selection Time~(ST): Time from target appearance to selection, relevant as follow-up manipulations can impact selection times.
    \item Drag~\&~Drop Time~(DDT): The time from selection to dropping. 
    \item Error Rate~(ER): The trial if the target’s center is not within the target’s radius.
    \item Subjective Task Load: NASA-TLX (Task Load Index) questionnaire~\cite{nasatlx} to assess subjective workload and questions about eye and hand fatigue referred to previous work~\cite{wagner_eye-hand_2024}.
\end{itemize}


\section{Result}
For the quantitative analysis, we first removed outliers using the interquartile range~(IQR). We detected that a task completion time exceeds the mean + 3 × SD or tracking-induced errors. In total, we removed 130 out of 3,960 trials ~(3.28\%). Then, we checked the normality of the data and applied a log transformation for non-normally distributed factors. We used repeated-measures ANOVA with Greenhouse-Geisser corrections when sphericity was violated. We also performed pairwise comparisons with Bonferroni correction to adjust the p-value. For subjective data, we used a Friedman test with Bonferroni corrected post-hoc Conover tests.

\begin{figure}[t]
    \centering
    \includegraphics[width=0.95\columnwidth]{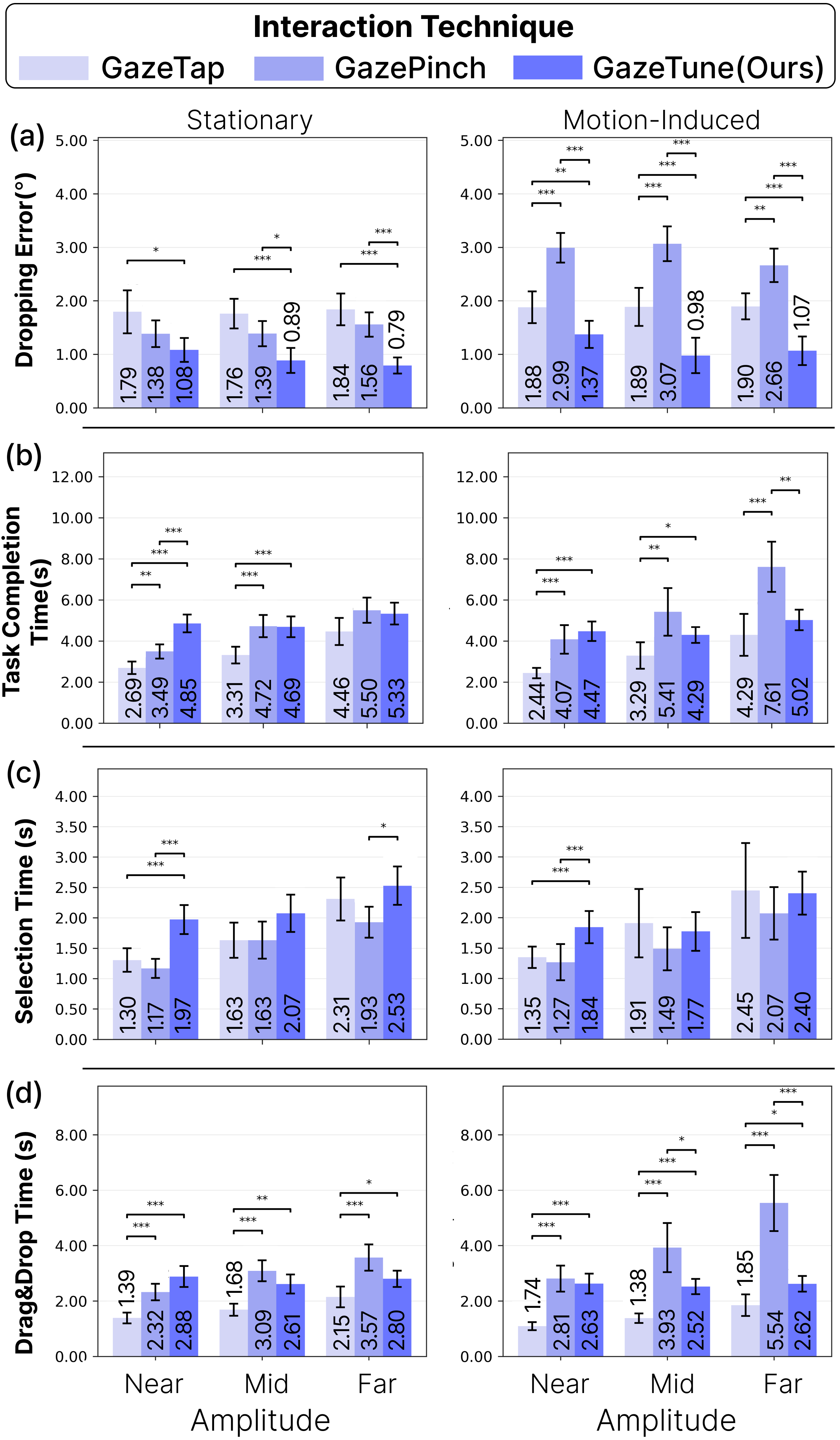}
    \caption{Results across context and amplitude for each technique with the main effect of technique.}
    \Description{Amplitude-wise performance comparison across contexts. The figure expands the same four dependent measures by separating results into Near, Mid, and Far amplitudes for \textit{Stationary} and \textit{Motion-Induced} conditions. Each panel shows grouped bars for GazeTap, GazePinch, and GazeTune, allowing comparison of how amplitude changes the relative performance of the three techniques. Larger amplitudes generally make the task more difficult, especially in the \textit{Motion-Induced} condition, and this increase is most pronounced for GazePinch. GazeTune remains comparatively stable across amplitudes, particularly in dropping error, suggesting that its refinement mechanism scales better as movement distance increases. The figure helps the reader see not only overall technique differences but also where those differences become larger.}
    \label{fig:performance2}
\end{figure}

\subsection{Performance}

\subsubsection{Dropping Error (DE)}
We observed main effects of \textsc{context} ($F_{19.00}^{1.00}=42.47, p<.001$), \textsc{technique}~($F_{36.21}^{1.91}=64.52, p<.001$), and \textsc{context} × \textsc{technique}~($F_{28.01}^{1.47}=32.90, p<.001$). Errors increased under \textit{Motion-Induced} overall, but the pattern was technique-specific. \textit{GazePinch} showed a marked rise (1.44$\rightarrow$2.91, $\Delta{=}+1.47$, $p{<}.001$), \textit{GazeTune} a smaller rise (0.92$\rightarrow$1.14, $\Delta{=}+0.22$, $p{<}.05$), whereas \textit{GazeTap} was relatively stable (1.80$\rightarrow$1.89, $\Delta{=}+0.09$).
In \textit{Stationary}, \textsc{technique} was significant~($F_{33.87}^{1.78}=25.24, p<.001$), while \textsc{amplitude} and the \textsc{technique} × \textsc{amplitude} were not significant. \textit{GazeTune} minimized error across amplitudes.
In \textit{Motion-Induced}, \textsc{technique}~($F_{32.45}^{1.71}=66.72, p<.001$), \textsc{amplitude}~($F_{34.67}^{1.82}=4.86, p=.016$), and \textsc{technique} × \textsc{amplitude}~($F_{32.91}^{1.73}=3.71, p=.041$) were all significant.
\textit{GazeTune} again had the lowest error, and \textit{GazeTap} was lower than \textit{GazePinch}.
\textit{Motion-Induced} increased errors and widened technique gaps, with \textit{GazeTune} lower than \textit{GazeTap} and \textit{GazePinch}.

\subsubsection{Task Completion Time (TCT)}
The result showed a main effect of \textsc{technique}~($F_{26.60}^{1.40}=25.68, p<.001$), but no main effect of \textsc{context}. An \textsc{context} × \textsc{technique} was also observed~($F_{15.82}^{0.83}=16.03, p<.002$). Technique ordering stayed the same, but the gap differed by context.
\textit{GazePinch} slowed under \textit{Motion-Induced} (4.57$\rightarrow$5.70\,s), whereas \textit{GazeTap} and \textit{GazeTune} showed no reliable change.
In \textit{Stationary}, both \textsc{technique}~($F_{31.38}^{1.65}=23.33, p<.001$) and \textsc{amplitude}~($F_{30.99}^{1.63}=57.73, p<.001$) showed significant main effects with a significant interaction~($F_{38.72}^{2.04}=7.93, p=.001$). 
Participants were fastest with \textit{GazeTap}, and the time worsened with amplitude.
The technique gap changed across amplitudes.
In \textit{Motion-Induced}, the same pattern held that \textit{GazeTap} remained fastest, and amplitude led to the slower time. 
Also, the \textsc{technique}~($F_{26.96}^{1.42}=22.97, p<.001$), \textsc{amplitude}~($F_{33.84}^{1.78}=46.91, p<.001$) and \textsc{Technique} × \textsc{amplitude} interaction were significant~($F_{29.24}^{1.54}=11.62, p<.001$).

\subsubsection{Selection Time (ST)}
Within Amplitude, we found a main effect of \textsc{technique}~($F_{30.89}^{1.62}=12.96,\ p<.001$), but no main effect of \textsc{context} and \textsc{context} × \textsc{technique} were observed. Descriptively, \textit{GazePinch} yielded the shortest selection times~(1.58$\rightarrow$1.61\,s), followed by \textit{GazeTap} (1.75$\rightarrow$1.90\,s) and \textit{GazeTune} (2.19$\rightarrow$2.01\,s).

In \textit{Stationary}, the effect of \textsc{technique}~($F_{27.06}^{1.42}=13.64, p<.001$), \textsc{amplitude}~($F_{34.31}^{1.81}=36.24, p<.001$) and \textsc{technique} × \textsc{amplitude}~($F_{34.27}^{1.80}=3.63, p=.041$) were significant. \textit{GazeTune} was reliably slower than \textit{GazeTap} and \textit{GazePinch}, and amplitude increased time. 
In \textit{Motion-Induced}, \textsc{technique}~($F_{37.44}^{1.97}=11.64, p<.001$ and \textsc{amplitude}~($F_{32.66}^{1.72}=56.52, p<.001$) were significant but the \textsc{technique} × \textsc{amplitude} was not. The result shows \textit{GazeTune} was slower, and the amplitude increased the selection time. \textit{GazePinch} and \textit{GazeTap} was faster than \textit{GazeTune} in both contexts.

\subsubsection{Drag\&Drop Time (DDT)}
Within Amplitude, there was a main effect of \textsc{technique}~($F_{22.67}^{1.19}=46.10, p<.001$), and a \textsc{context} × \textsc{technique}~($F_{13.37}^{0.70}=20.13, p=.001$). The \textsc{context} was not significant.
It showed that \textit{Motion-Induced} reduced DDT for \textit{GazeTap} (1.74$\rightarrow$1.44\,s, $\Delta{=}-0.30$\,s, $p{<}.01$) but increased it for \textit{GazePinch} (2.99$\rightarrow$4.09\,s, $\Delta{=}+1.10$\,s, $p{<}.001$), with no reliable change for \textit{GazeTune}~(2.77$\rightarrow$2.59\,s).
\textit{GazeTap} was faster than \textit{GazePinch} and \textit{GazeTune} in both contexts, with \textit{Motion-Induced} sometimes favoring \textit{GazeTune} over \textit{GazePinch}.
In \textit{Stationary}, \textsc{technique}~($F_{24.05}^{1.27}=30.86, p<.001$) and \textsc{amplitude}~($F_{34.23}^{1.80}=24.60, p<.001$) were significant. \textsc{technique} × \textsc{amplitude} was also significant~($F_{32.01}^{1.68}=8.49, p=.002$).
\textit{GazeTap} was the fastest across amplitudes, and amplitudes make it slower. 
In \textit{Motion-Induced}, \textsc{technique}~($F_{21.95}^{1.16}=43.62, p<.001$), \textsc{amplitude}~($F_{32.09}^{1.69}=30.41, p<.001$), \textsc{technique} × \textsc{amplitude}~($F_{21.99}^{1.16}=18.67, p<.001$) were significant.
\textit{GazeTune} sometimes outperformed \textit{GazePinch}.

\subsubsection{Error Rate (ER)}
In \textit{Stationary}, the ordering is \textit{GazeTune}~(2.88\%) < \textit{GazePinch}~(5.91\%) < \textit{GazeTap}~(8.94\%). 
Under \textit{Motion-Induced}, the pattern was different~(\textit{GazeTune}~(4.24\%) < \textit{GazeTap}~(9.70\%) < \textit{GazePinch} (37.12\%)) and widens markedly due to a large increase for \textit{GazePinch} while \textit{GazeTap} is highest in the seated context. \textit{GazePinch} exhibits a substantial rise with movement, whereas \textit{GazeTune} remains comparatively low with a smaller increase and dispersion.

\begin{table}[t]
\centering
\caption{Drop-error rate by Context $\times$ Technique (DE$>$3.5°).}
\label{tab:ctx_tech_mean_sd_ci}
\begin{tabular}{llccc}
\toprule
\textbf{Context} & \textbf{Technique} & \textbf{Mean (\%)} & \textbf{SD (\%)} \\
\midrule
Stationary & GazeTap & 8.94\% & 12.65\% \\
Stationary & GazePinch & 5.91\% & 6.09\% \\
Stationary & GazeTune & \textbf{2.88\%} & \textbf{2.86\%} \\
Motion-Induced & GazeTap & 9.70\% & 14.76\% \\
Motion-Induced & GazePinch & 37.12\% & 15.58\%\\
Motion-Induced & GazeTune & \textbf{4.24\%} & \textbf{4.85\%} \\
\bottomrule
\end{tabular}
\end{table}
\setlength{\textfloatsep}{6pt} 

\begin{figure}[t]
    \centering
    \includegraphics[width=0.99\columnwidth]{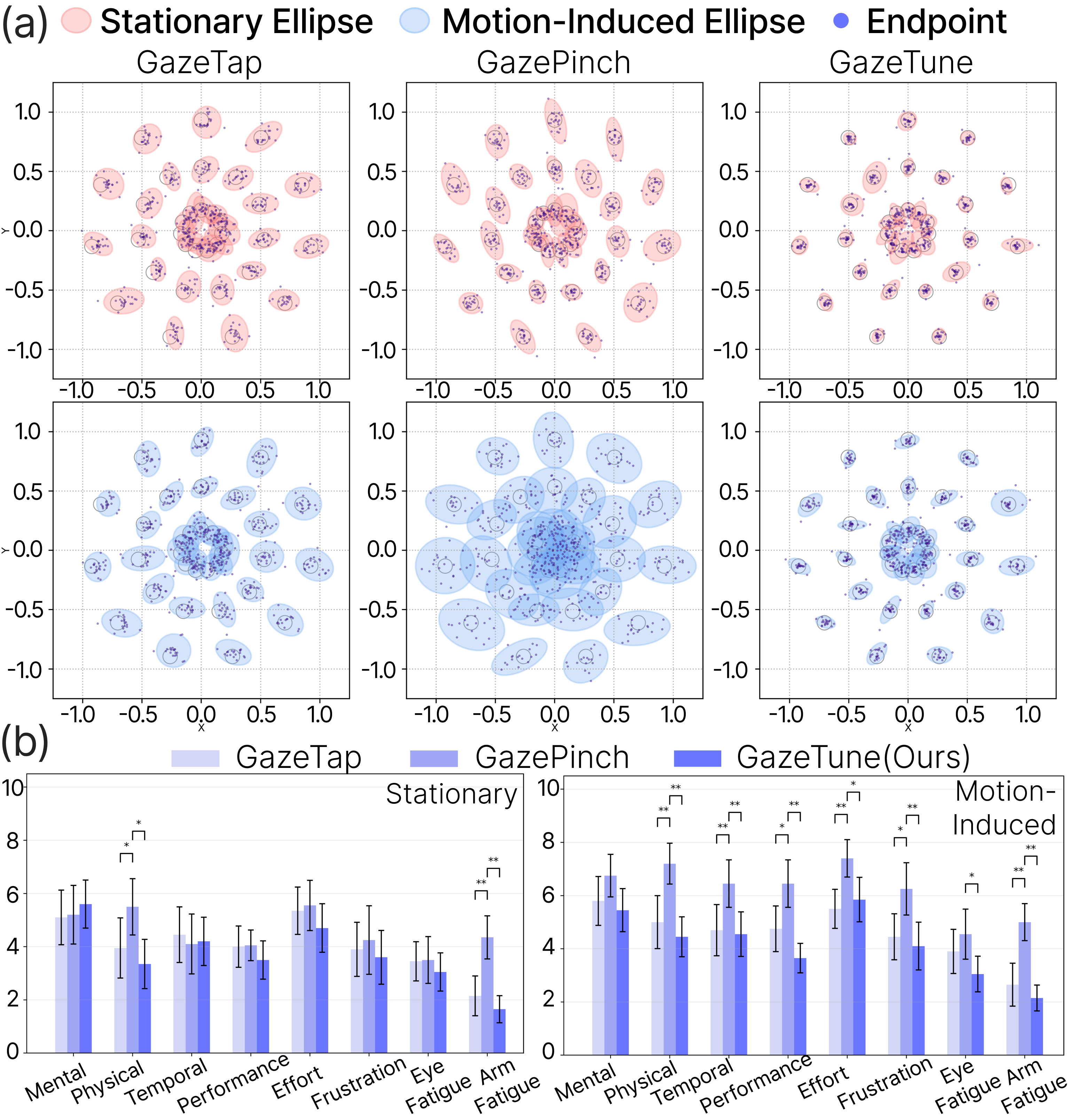}
    \caption{(a)~Release point distribution of each technique on each target as a 95\% confidence ellipse in visual angle. The removed outlier endpoint was excluded for visualization. (b)~NASA-TLX and subject rating of each technique for \textit{Stationary} and \textit{Motion-Induced} conditions.}
    \Description{Endpoint distributions across techniques and motion conditions. The figure compares GazeTap, GazePinch, and GazeTune using endpoint scatterplots with ellipse summaries in \textit{Stationary} and \textit{Motion-Induced} contexts. Each technique occupies one column, and the two contexts are shown as separate rows, allowing direct comparison of endpoint spread and concentration. GazeTune shows the most compact endpoint clusters in both contexts, while GazePinch becomes much more dispersed under motion. The lower part of the figure adds subjective ratings for workload and fatigue measures, linking spatial variability to user experience. The main visual pattern is that GazeTune remains tighter and less affected by motion than the other two techniques.}
    \label{fig:drop endpoint}
\end{figure}

\subsection{Release point for Dragged Target}
\label{tab:releasedpoint}
\begin{table}[t]
\centering
\small
\caption{Mean ± SD of the 95\% ellipse size in visual angle across amplitudes, techniques, and contexts.}
\resizebox{\columnwidth}{!}{
\begin{tabular}{llcccc}
\toprule
\multirow{2}{*}{Amplitude} & \multirow{2}{*}{Technique} & \multicolumn{2}{c}{Stationary} & \multicolumn{2}{c}{Motion-Induced} \\
\cmidrule(lr){3-4}\cmidrule(lr){5-6}
 &  & $\overline{x}\ (\pm\ s_x)$ (\degree) & $\overline{y}\ (\pm\ s_y)$ (\degree) & $\overline{x}\ (\pm\ s_x)$ (\degree) & $\overline{y}\ (\pm\ s_y)$ (\degree) \\
\midrule
\multirow{3}{*}{Near} & GazeTap & 7.32 $\pm$ 1.22 & 4.70 $\pm$ 0.70 & 7.65 $\pm$ 1.29 & 5.34 $\pm$ 0.98 \\
 & GazePinch & 6.97 $\pm$ 1.67 & 4.22 $\pm$ 1.09 & 12.68 $\pm$ 1.36 & 9.96 $\pm$ 1.42 \\
 & GazeTune & 6.11 $\pm$ 1.59 & 3.23 $\pm$ 0.87 & 6.99 $\pm$ 0.94 & 4.63 $\pm$ 1.04 \\
\addlinespace
\multirow{3}{*}{Mid} & GazeTap & 7.29 $\pm$ 0.77 & 4.91 $\pm$ 0.75 & 8.28 $\pm$ 1.04 & 5.34 $\pm$ 0.98 \\
 & GazePinch & 7.26 $\pm$ 1.62 & 4.44 $\pm$ 0.81 & 13.11 $\pm$ 0.95 & 10.26 $\pm$ 1.29 \\
 & GazeTune & 4.97 $\pm$ 1.42 & 3.43 $\pm$ 1.01 & 5.96 $\pm$ 0.97 & 3.07 $\pm$ 0.57 \\
\addlinespace
\multirow{3}{*}{Far} & GazeTap & 8.37 $\pm$ 0.94 & 5.80 $\pm$ 0.90 & 8.60 $\pm$ 1.19 & 6.44 $\pm$ 1.05 \\
 & GazePinch & 9.03 $\pm$ 1.52 & 5.21 $\pm$ 1.22 & 13.83 $\pm$ 1.65 & 9.76 $\pm$ 1.78 \\
 & GazeTune & 4.41 $\pm$ 1.21 & 3.13 $\pm$ 0.50 & 6.25 $\pm$ 1.36 & 3.83 $\pm$ 1.17 \\
\addlinespace
\bottomrule
\end{tabular}
}
\end{table}

During drag \& drop, the final release is the most vulnerable to motion. To characterize spatial accuracy, we observed the release point distribution with 95\% confidence ellipses, which reveals mobility-driven dispersion and whether anchored refinement steadies release. Detailed values are reported in table~\ref{tab:releasedpoint}.
First, ellipses are generally larger under \textit{Motion-Induced}, with the gap most pronounced for \textit{GazePinch}. 
For example, \textit{GazePinch} at Mid is $13.11^{\circ}\times 10.26^{\circ}$ when \textit{Motion-Induced} vs.\ $7.26^{\circ}\times 4.44^{\circ}$ under \textit{Stationary}. 
The difference persists at Far~(\textit{Motion-Induced} $13.83^{\circ}\times 9.76^{\circ}$ vs.\ \textit{Stationary}: $9.03^{\circ}\times 5.21^{\circ}$).
Moreover, GazePinch is the largest, and GazeTune is the smallest.
Under Motion-Induced, \textit{GazePinch} yields the largest dispersion at all amplitudes, while \textit{GazeTune} is consistently the smallest. 
The same ordering holds for \textit{Stationary} (e.g., \textit{GazeTune} at Far $4.41^{\circ}\times 3.13^{\circ}$).
In addition, ellipse sizes generally increase toward Far. For \textit{GazeTap}, ellipse size grows from Near$\rightarrow$Mid$\rightarrow$Far in both contexts. 
\textit{GazePinch} increases horizontally with amplitude under \textit{Motion-Induced} while its vertical extent peaks at Mid.
Finally, we consistently observe that horizontal exceeds vertical and SD varies. In \textit{Motion-Induced}-\textit{GazePinch} at Far aspect ratio was 1.42 (13.83/9.76).
In \textit{Stationary}-\textit{GazeTune} at Near, it was 1.89 (6.11/3.23). 
This indicates a stronger horizontal error component. Regarding the SD, \textit{Motion-Induced} with GazePinch shows the largest variability, whereas \textit{GazeTune} under \textit{Stationary} is comparatively stable.

\subsection{Task Load}
For \textit{Stationary}, we observed differences mainly in physical load. \textit{GazePinch} is perceived as more demanding on physical demand and arm fatigue, while other factors are comparable. Physical demand differed by technique ($\chi^2$(2)=15.44, p<.001, W=.39): means were \textit{GazeTune}=3.35 < \textit{GazeTap}=3.95 < \textit{GazePinch}=5.50. Post-hoc tests indicated \textit{GazeTap} < \textit{GazePinch} (p=.011) and \textit{GazeTune} < \textit{GazePinch} (p=.016), with no difference between \textit{GazeTap} and \textit{GazeTune}. Arm fatigue showed a strong effect ($\chi^2$(2)=27.38, p<.001, W=.68) with \textit{GazeTune}=1.65 < \textit{GazeTap}~(2.15) < \textit{GazePinch}~(4.35) and post-hoc revealed \textit{GazeTap} < \textit{GazePinch} (p=.003) and \textit{GazeTune} < \textit{GazePinch} (p<.001), while \textit{GazeTap} vs. \textit{GazeTune} was not significant.

For \textit{\textit{Motion-Induced}}, the effects of different techniques had a major impact, reaching significance for \textit{Physical demand} ($\chi^2(2)=25.14$, $p<.001$, $W=.63$), \textit{Temporal demand} ($\chi^2(2)=18.38$, $p<.001$, $W=.46$), \textit{Performance} ($\chi^2(2)=17.22$, $p<.001$, $W=.43$), \textit{Effort} ($\chi^2(2)=13.00$, $p=.002$, $W=.33$), \textit{Frustration} ($\chi^2(2)=19.54$, $p<.001$, $W=.49$), \textit{Eye Fatigue} ($\chi^2(2)=10.78$, $p=.005$, $W=.27$), \textit{Arm Fatigue} ($\chi^2(2)=25.49$, $p<.001$, $W=.64$) while \textit{Mental demand} was not significant. During \textit{Motion-Induced}, technique differences broaden and intensify. A consistent pattern across Effort/Frustration/Physical/Temporal is \textit{GazeTap} < \textit{GazePinch} < \textit{GazeTune}~(lower workload for \textit{GazeTap}, highest for \textit{GazeTune}), while Performance favors \textit{GazeTune} best.

\subsection{Summary}
In this section, we summarize the key findings for each metric.
Details of the user feedback are provided in the Appendix~\ref{a:userfeedback}.
For Dropping Error, \textit{GazeTune} consistently reduced errors relative to \textit{GazeTap} and \textit{GazePinch} across contexts and amplitudes.
While \textit{GazePinch} enabled more accurate drops than \textit{GazeTap} in \textit{Stationary}, \textit{GazePinch} showed a higher error angle in \textit{Motion-Induced}.
In comparison, \textit{GazeTune} was more resilient than arm movement dragging in mobile conditions, confirming the achievements of our design goals~(precision under mobility).
About Task Completion Time, \textit{GazeTap} was the fastest overall, while \textit{GazeTune} became comparable to \textit{GazePinch} under \textit{Motion-Induced} without large time penalties in mobile use.
For shorter amplitudes, \textit{GazeTap} was more advantageous, but the difference diminished as the amplitude increased.
Regarding Selection Time, \textit{GazeTune} was slower than \textit{GazePinch} at shorter amplitudes.
Because \textit{GazeTune} does not require gaze to be precisely within the target, users often place gaze coarsely near the target and then perform additional confirmation drag with dwell phase.
For Drag\&Drop Time, \textit{GazeTap} was faster than both \textit{GazePinch} and \textit{GazeTune} in \textit{Stationary} and \textit{Motion-Induced} conditions.
Notably, \textit{GazeTune} outperformed \textit{GazePinch} in the \textit{Motion-Induced} condition.
In Error Rate, \textit{GazeTune} shows the lower error rate across contexts.
\textit{GazePinch} can be acceptable in a \textit{Stationary} environment, yet it becomes risky in \textit{Motion-Induced} tasks where drop errors surge.




\begin{figure}[t]
    \centering
    \includegraphics[width=0.98\linewidth]{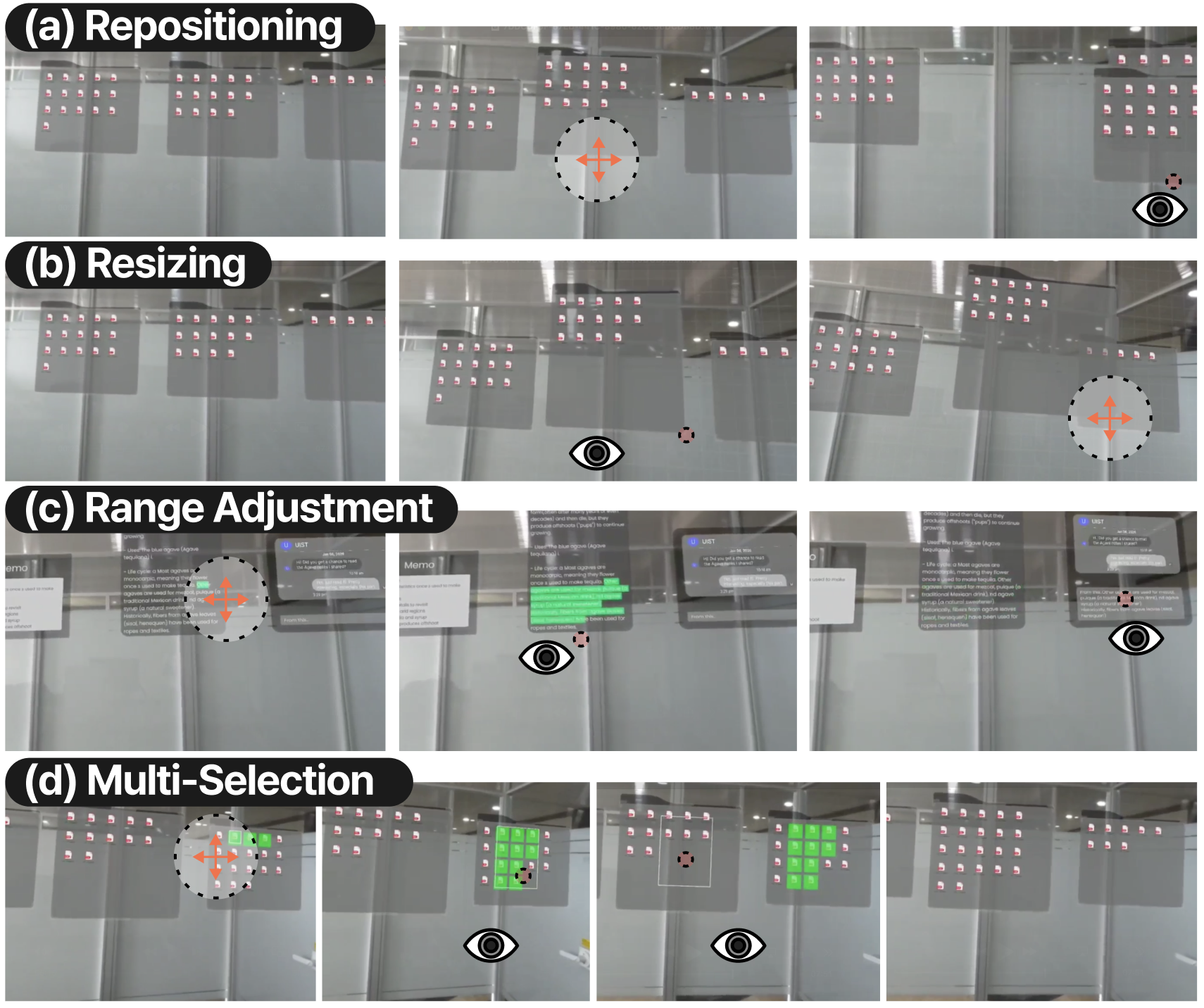}
    \Description{Example interaction scenarios supported by the technique. Four rows illustrate different interface tasks using floating interface elements: repositioning, resizing, range adjustment, and multi-selection. Each row shows a short first-person sequence in which the user first acquires or moves toward the target coarsely, then performs local refinement inside a lock region. The manipulated object differs across rows, but the interaction principle remains the same. In repositioning, a window is moved to a new location; in resizing, the user adjusts a resize handle; in range adjustment, a selected span is extended or corrected; and in multi-selection, a larger region is defined to include multiple items. The figure shows that the same interaction method extends beyond simple pointing to several common interface operations.}
    \caption{Use case scenarios of \textit{GazeTune}.}
    \label{fig:example}
\end{figure}

\section{Use Case Scenario}
We illustrate how gaze-and-touch supports a unified interaction pattern across near-far tasks that require both rapid navigation and precise control in productivity-oriented AR~(Figure~\ref{fig:example}).
We further argue that the technique supports near-far interaction by enabling seamless transitions between refinement within the gaze scope and rapid repositioning outside the scope.
In other words, users can (1) acquire a target or region quickly at a distance, (2) snap the interaction focus to the intended area, and (3) perform stable, fine-grained adjustments within gaze scope.
This consistency allows the workflow to generalize across tasks that repeatedly alternate between coarse navigation and precise adjustment.

\textbf{Repositioning.}
Users first acquire the target to move with gaze and anchor refinement. 
While the gaze remains within the scope, they perform small nudges to align the object locally. 
When the destination lies farther away, they shift their gaze outside the scope to trigger snapping and then re-enter the scope to refine the final placement with touch.

\textbf{Resizing.}
Users gaze at and select the resize handle and create an anchor for refinement. 
Within the scope, refinement supports fine-scale tuning (e.g., matching an edge or fitting a layout).
To reach a substantially different size, users look outside the scope to snap the scale toward a new range quickly, then return within the scope for precise final adjustment.

\textbf{Range adjustment.}
Users first anchor the start point and refine it within the scope for an exact boundary.
To adjust the extent over a larger distance, they shift gaze outside the scope to snap toward a distant candidate endpoint.
They then return within the scope to refine the endpoint precisely, enabling iterative tuning by alternating between snapping and local refinement.

\textbf{Multi-selection.}
Users anchor an initial item or a boundary point and refine the selection locally within the scope.
They then expand coverage by moving the gaze outside the scope to snap the selection window across a larger region and re-enter the scope to refine the boundary again.
This supports both precise picking and rapid area coverage.




\section{Discussion}
Our findings show that combining rapid gaze acquisition with touch-based refinement enables accurate and reliable dragging while maintaining competitive interaction time.
A key strength of our approach is the enhanced input stability, achieved by anchoring the cursor to the gaze position and maintaining continuous touch contact.
By anchoring the cursor, the user could reduce demanded attention and enable finer and more controlled adjustments. 
In contrast, mid-air manipulation requires sustained arm elevation, which is unstable and even causes the ``Gorilla Arm'' effect~\cite{jang2017modeling}. 
These observations indicate that combining gaze acquisition with touch refinement could effectively balance speed and precision in gaze-based interaction.

\textbf{Balancing Efficiency and Robustness in Mobile Gaze-Driven Interaction~(RQ1).}
While \textit{GazeTune} reduced dropping error, it introduced a trade-off in initial ST, particularly for near targets.
This delay inherently stems from the interaction design, which pairs gaze anchoring and touch refinement with a 0.3s dwell confirmation.
We observed that even for easy targets~(3.5°), users naturally preferred a coarse glance followed by touch adjustment rather than relying on gaze alone.
However, this slight increase in ST is outweighed by a reduction in DDT, especially under \textit{Motion-Induced} instability where baseline methods suffered from repeated errors.
To mitigate the initial selection delay, our system is designed to integrate the quick-selection capability of \textit{GazeTap}.
Users can fluidly bypass touch refinement when high precision is unnecessary.
Under \textit{Motion-Induced} instability and at longer viewing amplitudes, \textit{GazeTune} provides more reliable accuracy while maintaining comparable completion times. 
This motivates a progressive engagement strategy that begins with rapid gaze targeting and invokes touch refinement when error risk is high~(e.g., peripheral placements).
Ultimately, this approach benefits spatial computing in large workspaces by maintaining a single, fluid workflow with localized adjustments rather than demanding mode switching.

\textbf{Cascaded Synergy of Gaze and Touch~(RQ2).}
The consistent performance of \textit{GazeTune} under both \textit{Stationary} and \textit{Motion-Induced} conditions highlights its robust control capabilities. 
By anchoring the cursor before refinement and re-anchoring for larger repositioning, the interaction reduces reliance on prolonged gaze fixation or sustained mid-air arm movement and ensures stability during body motion.
From a motor-control perspective, our cascaded interaction structure includes the ballistic and homing phases of gaze-driven pointing, where touch-based refinement stabilizes the homing phase.
This likely helps \textit{GazeTune} maintain comparable performance across near and far targets even under \textit{Motion-Induced} conditions.
Furthermore, we found that touch-based refinement offers more practical benefits during continuous post-selection manipulation than during initial target acquisition. The integration of gaze and touch is most effective when their roles are complementary: gaze supports rapid coarse relocation, and touch enables localized stabilization and correction.
Together, these properties suggest that \textit{GazeTune} is well-suited for gaze interaction in \textit{Motion-Induced} contexts with a natural gaze-driven workflow.

\textbf{Behavioral Strategies and Experience of GazeTune~(RQ3).}
Participants adapted quickly to \textit{GazeTune}, although it was initially less intuitive than \textit{GazeTap} and \textit{GazePinch}.
This was primarily due to a temporary learning curve associated with the visual split-attention and spatial-mapping coordination required to look at a target in 3D while moving a finger blindly across a 2D wrist-worn interface.
However, NASA-TLX results showed no significant negative impact on cognitive demand.
As participants adapted easily after a few trials, the initial attentional demand rapidly diminished.
Noticeable advantages were observed at the far-edge targets, where small finger adjustments provided more stable placement than head realignment.
Interestingly, the average time for each measure in the \textit{Motion-Induced} condition was even slightly faster than in the \textit{Stationary} condition.  
Users tended to stop refining once the anchored refinement cursor reached a sufficiently accurate region, rather than pursuing marginal improvement.
In other words, the touch refinement helps them reach quickly toward the target region, but they simply do not overcorrect.
Motion also introduces attention and safety considerations that discourage aggressive micro-corrections, leading participants to end sequences earlier.

\textbf{Limitations \& Future Work.}
For our evaluation, we limited the mobility condition to ordinary walking speed, as continuous manipulation is rarely required at running speeds.
Although evaluation on a treadmill does not fully reproduce the visual motion cues, we intentionally chose this setup to isolate continuous gait-induced noise without self-regulated walking speed.
To strengthen generalizability, future work should examine real-world walking scenarios (e.g., circuit-style paths~\cite{shin_using_2025}) or outdoor environments with changing visual backdrop during the movement.
Also, our current evaluation lacks a detailed temporal breakdown of the interaction phases. While a DDT comparison with GazeTap (which lacks a refinement phase) provides a rough estimate of one second or less for GazeTune’s manual refinement, this is merely an approximation.
Future work should conduct a finer-grained temporal analysis to better understand the temporal overhead of the refinement process.
Additionally, our experimental procedure used fixed blocks, so the mobility condition's order effect was not considered.
We also fixed the UI depth at 2~m, which reflects a common spatial UI, but spatial computing systems deploy targets across varied depths.
Our results do not guarantee depth-invariant performance. 
A promising direction is to adapt the gain of refinement to viewing amplitude as depth increases.
Additionally, our evaluation focused on a path-anchored interface, whereas spatial interfaces can be anchored to different reference frames~(e.g., world-fixed or object-attached).
Future work should investigate the refinement behavior under diverse anchoring conditions.
In a few cases, users initiated refinement near the edge of the smartwatch, which is a limitation that requires restarting and constrains further adjustment. 
While our single interaction sequence allowed users to reach most intended positions, edge-case robustness could be further improved by supporting a brief clutching period as an additional gaze-touch gating mechanism that suspends touch until a new gaze anchor is established, allowing safe finger repositioning.
Another interesting direction would be to explore surface-based interaction beyond device-bound touch surfaces. For example, on-skin input on the hand or upper arm could provide a larger interaction area with the benefits of surface contact, particularly as sensing technologies continue to improve.
Finally, participants suggested system-level enhancements, including increasing the refinement resolution (P2) and adding additional interaction layers or dimensions (e.g., press-based refinements; P19). These extensions could offer richer control while retaining the fluid gaze-driven workflow.

%
\vspace{-0.35\baselineskip}

\section{Conclusion}
In this paper, we introduce \textit{GazeTune}, a cascaded gaze-and-touch technique that supports precise and stable gaze-driven interaction in XR.
By offloading coarse acquisition to gaze and limiting touch to anchored refinement, \textit{GazeTune} preserves fast gaze pointing while improving precision. 
Furthermore, our cascaded design enables interaction within a single continuous sequence to minimize clutching and reduce reliance on prolonged visual fixation.
To validate the effectiveness of \textit{GazeTune}, we conducted a comparative study against gaze-only and gaze-pinch baselines.
The results indicate that \textit{GazeTune} achieved significantly lower error and maintained consistent dropping accuracy even under \textit{Motion-Induced} conditions with comparable task completion time.
In summary, touch input serves as a selective trigger for minimal physical effort, effectively functioning as a precise control interface for selection and drag~\&~drop under physical mobility.
We believe our study highlights the potential of cascaded and state-aware input to enable fluid and precise interaction.

\begin{acks}
This work was partly supported by the National Research Foundation of Korea(NRF) grant (RS-2026-25470200, 40\%), Institute of Information \& communications Technology Planning \& Evaluation(IITP) \& ITRC(Information Technology Research Center) grant (IITP-2026-RS-2024-00436398, 50\%), and the IITP under the virtual convergence support program to nurture the best talents(IITP-2022(2026)-RS-2022-00156435, 10\%) grant funded by the Korea government(MSIT).


\end{acks}

\bibliographystyle{ACM-Reference-Format}
\bibliography{Main}

@inproceedings{ref:Toward_everyday_gaze,
author = {Feit, Anna Maria and Williams, Shane and Toledo, Arturo and Paradiso, Ann and Kulkarni, Harish and Kane, Shaun and Morris, Meredith Ringel},
title = {Toward Everyday Gaze Input: Accuracy and Precision of Eye Tracking and Implications for Design},
year = {2017},
isbn = {9781450346559},
publisher = {Association for Computing Machinery},
address = {New York, NY, USA},
url = {https://doi.org/10.1145/3025453.3025599},
doi = {10.1145/3025453.3025599},
booktitle = {Proceedings of the 2017 CHI Conference on Human Factors in Computing Systems},
pages = {1118–1130},
numpages = {13},
location = {Denver, Colorado, USA},
series = {CHI '17}
}

@inproceedings{wagner_eye-hand_2024,
	address = {Pittsburgh PA USA},
	title = {Eye-{Hand} {Movement} of {Objects} in {Near} {Space} {Extended} {Reality}},
	isbn = {979-8-4007-0628-8},
	url = {https://dl.acm.org/doi/10.1145/3654777.3676446},
	doi = {10.1145/3654777.3676446},
	language = {en},
	urldate = {2025-09-01},
	booktitle = {Proceedings of the 37th {Annual} {ACM} {Symposium} on {User} {Interface} {Software} and {Technology}},
	publisher = {ACM},
	author = {Wagner, Uta and Asferg Jacobsen, Andreas and Feuchtner, Tiare and Gellersen, Hans and Pfeuffer, Ken},
	month = oct,
	year = {2024},
	pages = {1--13}
}

@inproceedings{khamis_eyescout_2017,
	address = {Québec City QC Canada},
	title = {{EyeScout}: {Active} {Eye} {Tracking} for {Position} and {Movement} {Independent} {Gaze} {Interaction} with {Large} {Public} {Displays}},
	isbn = {978-1-4503-4981-9},
	shorttitle = {{EyeScout}},
	url = {https://dl.acm.org/doi/10.1145/3126594.3126630},
	doi = {10.1145/3126594.3126630},
	language = {en},
	urldate = {2025-09-01},
	booktitle = {Proceedings of the 30th {Annual} {ACM} {Symposium} on {User} {Interface} {Software} and {Technology}},
	publisher = {ACM},
	author = {Khamis, Mohamed and Hoesl, Axel and Klimczak, Alexander and Reiss, Martin and Alt, Florian and Bulling, Andreas},
	month = oct,
	year = {2017},
	pages = {155--166}
}

@inproceedings{chatterjee_gazegesture_2015,
	address = {New York, NY, USA},
	series = {{ICMI} '15},
	title = {Gaze+{Gesture}: {Expressive}, {Precise} and {Targeted} {Free}-{Space} {Interactions}},
	isbn = {978-1-4503-3912-4},
	shorttitle = {Gaze+{Gesture}},
	url = {https://dl.acm.org/doi/10.1145/2818346.2820752},
	doi = {10.1145/2818346.2820752},
	urldate = {2025-08-31},
	booktitle = {Proceedings of the 2015 {ACM} on {International} {Conference} on {Multimodal} {Interaction}},
	publisher = {Association for Computing Machinery},
	author = {Chatterjee, Ishan and Xiao, Robert and Harrison, Chris},
	month = nov,
	year = {2015},
	pages = {131--138}
}

@article{deng_understanding_2017,
	title = {Understanding the impact of multimodal interaction using gaze informed mid-air gesture control in {3D} virtual objects manipulation},
	volume = {105},
	issn = {1071-5819},
	url = {https://www.sciencedirect.com/science/article/pii/S1071581917300629},
	doi = {10.1016/j.ijhcs.2017.04.002},
	urldate = {2025-09-01},
	journal = {International Journal of Human-Computer Studies},
	author = {Deng, Shujie and Jiang, Nan and Chang, Jian and Guo, Shihui and Zhang, Jian J.},
	month = sep,
	year = {2017},
	pages = {68--80}
}

@inproceedings{yu_gaze-supported_2021,
	address = {New York, NY, USA},
	series = {{CHI} '21},
	title = {Gaze-{Supported} {3D} {Object} {Manipulation} in {Virtual} {Reality}},
	isbn = {978-1-4503-8096-6},
	url = {https://dl.acm.org/doi/10.1145/3411764.3445343},
	doi = {10.1145/3411764.3445343},
	urldate = {2025-08-31},
	booktitle = {Proceedings of the 2021 {CHI} {Conference} on {Human} {Factors} in {Computing} {Systems}},
	publisher = {Association for Computing Machinery},
	author = {Yu, Difeng and Lu, Xueshi and Shi, Rongkai and Liang, Hai-Ning and Dingler, Tilman and Velloso, Eduardo and Goncalves, Jorge},
	month = may,
	year = {2021},
	pages = {1--13}
}

@inproceedings{li_estimating_2025,
	address = {Yokohama Japan},
	title = {Estimating the {Effects} of {Encumbrance} and {Walking} on {Mixed} {Reality} {Interaction}},
	isbn = {979-8-4007-1394-1},
	url = {https://dl.acm.org/doi/10.1145/3706598.3713492},
	doi = {10.1145/3706598.3713492},
	language = {en},
	urldate = {2025-09-01},
	booktitle = {Proceedings of the 2025 {CHI} {Conference} on {Human} {Factors} in {Computing} {Systems}},
	publisher = {ACM},
	author = {Li, Tinghui and Velloso, Eduardo and Withana, Anusha and Sarsenbayeva, Zhanna},
	month = apr,
	year = {2025},
	pages = {1--24}
}

@inproceedings{palmeira2023quantifying,
  title={Quantifying the 'Gorilla Arm' Effect in a Virtual Reality Text Entry Task via Ray-Casting: A Preliminary Single-Subject Study},
  author={Palmeira, Eduardo GQ and Campos, Alexandre and Moraes, {\'I}gor A and de Siqueira, Alexandre G and Ferreira, Marcelo GG},
  booktitle={Proceedings of the 25th Symposium on Virtual and Augmented Reality},
  pages={274--278},
  year={2023}
}

@inproceedings{hsieh2016designing,
  title={Designing a willing-to-use-in-public hand gestural interaction technique for smart glasses},
  author={Hsieh, Yi-Ta and Jylh{\"a}, Antti and Orso, Valeria and Gamberini, Luciano and Jacucci, Giulio},
  booktitle={Proceedings of the 2016 CHI conference on human factors in computing systems},
  pages={4203--4215},
  year={2016}
}

@inproceedings{zhang_focusflow_2023,
	address = {New York, NY, USA},
	series = {{UIST} '23 {Adjunct}},
	title = {{FocusFlow}: {Leveraging} {Focal} {Depth} for {Gaze} {Interaction} in {Virtual} {Reality}},
	isbn = {979-8-4007-0096-5},
	shorttitle = {{FocusFlow}},
	url = {https://dl.acm.org/doi/10.1145/3586182.3615818},
	doi = {10.1145/3586182.3615818},
	urldate = {2025-08-31},
	booktitle = {Adjunct {Proceedings} of the 36th {Annual} {ACM} {Symposium} on {User} {Interface} {Software} and {Technology}},
	publisher = {Association for Computing Machinery},
	author = {Zhang, Chenyang and Chen, Tiansu and Nedungadi, Rohan Russel and Shaffer, Eric and Soltanaghai, Elahe},
	month = oct,
	year = {2023},
	pages = {1--4}
}

@inproceedings{chen_gazeraycursor_2023,
	address = {New York, NY, USA},
	series = {{VRST} '23},
	title = {{GazeRayCursor}: {Facilitating} {Virtual} {Reality} {Target} {Selection} by {Blending} {Gaze} and {Controller} {Raycasting}},
	isbn = {979-8-4007-0328-7},
	shorttitle = {{GazeRayCursor}},
	url = {https://dl.acm.org/doi/10.1145/3611659.3615693},
	doi = {10.1145/3611659.3615693},
	urldate = {2025-08-31},
	booktitle = {Proceedings of the 29th {ACM} {Symposium} on {Virtual} {Reality} {Software} and {Technology}},
	publisher = {Association for Computing Machinery},
	author = {Chen, Di Laura and Giordano, Marcello and Benko, Hrvoje and Grossman, Tovi and Santosa, Stephanie},
	month = oct,
	year = {2023},
	pages = {1--11}
}

@inproceedings{kumar_tagswipe_2020,
	address = {New York, NY, USA},
	series = {{CHI} '20},
	title = {{TAGSwipe}: {Touch} {Assisted} {Gaze} {Swipe} for {Text} {Entry}},
	isbn = {978-1-4503-6708-0},
	shorttitle = {{TAGSwipe}},
	url = {https://dl.acm.org/doi/10.1145/3313831.3376317},
	doi = {10.1145/3313831.3376317},
	urldate = {2025-08-31},
	booktitle = {Proceedings of the 2020 {CHI} {Conference} on {Human} {Factors} in {Computing} {Systems}},
	publisher = {Association for Computing Machinery},
	author = {Kumar, Chandan and Hedeshy, Ramin and MacKenzie, I. Scott and Staab, Steffen},
	month = apr,
	year = {2020},
	pages = {1--12}
}

@inproceedings{zhao_eyesaycorrect_2022,
	address = {Helsinki Finland},
	title = {{EyeSayCorrect}: {Eye} {Gaze} and {Voice} {Based} {Hands}-free {Text} {Correction} for {Mobile} {Devices}},
	isbn = {978-1-4503-9144-3},
	shorttitle = {{EyeSayCorrect}},
	url = {https://dl.acm.org/doi/10.1145/3490099.3511103},
	doi = {10.1145/3490099.3511103},
	language = {en},
	urldate = {2025-09-01},
	booktitle = {27th {International} {Conference} on {Intelligent} {User} {Interfaces}},
	publisher = {ACM},
	author = {Zhao, Maozheng and Huang, Henry and Li, Zhi and Liu, Rui and Cui, Wenzhe and Toshniwal, Kajal and Goel, Ananya and Wang, Andrew and Zhao, Xia and Rashidian, Sina and Baig, Furqan and Phi, Khiem and Zhai, Shumin and Ramakrishnan, Iv and Wang, Fusheng and Bi, Xiaojun},
	month = mar,
	year = {2022},
	pages = {470--482}
}

@inproceedings{brewster_multimodal_2003,
	address = {Ft. Lauderdale Florida USA},
	title = {Multimodal 'eyes-free' interaction techniques for wearable devices},
	isbn = {978-1-58113-630-2},
	url = {https://dl.acm.org/doi/10.1145/642611.642694},
	doi = {10.1145/642611.642694},
	language = {en},
	urldate = {2025-09-01},
	booktitle = {Proceedings of the {SIGCHI} {Conference} on {Human} {Factors} in {Computing} {Systems}},
	publisher = {ACM},
	author = {Brewster, Stephen and Lumsden, Joanna and Bell, Marek and Hall, Malcolm and Tasker, Stuart},
	month = apr,
	year = {2003},
	pages = {473--480}
}

@article{guarnera_automated_2025,
	title = {Automated {Fixation} {Error} {Correction} to {Support} {Eye} {Tracking} {Studies} on {Source} {Code}},
	volume = {9},
	issn = {2573-0142},
	url = {https://dl.acm.org/doi/10.1145/3725829},
	doi = {10.1145/3725829},
	language = {en},
	number = {3},
	urldate = {2025-09-01},
	journal = {Proc. ACM Hum.-Comput. Interact.},
	author = {Guarnera, Drew T. and Behler, Joshua A.C. and Sharif, Bonita and Maletic, Jonathan I.},
	month = may,
	year = {2025},
	pages = {1--17}
}

@inproceedings{sindhwani_retype_2019,
	address = {Glasgow Scotland Uk},
	title = {{ReType}: {Quick} {Text} {Editing} with {Keyboard} and {Gaze}},
	isbn = {978-1-4503-5970-2},
	shorttitle = {{ReType}},
	url = {https://dl.acm.org/doi/10.1145/3290605.3300433},
	doi = {10.1145/3290605.3300433},
	language = {en},
	urldate = {2025-09-01},
	booktitle = {Proceedings of the 2019 {CHI} {Conference} on {Human} {Factors} in {Computing} {Systems}},
	publisher = {ACM},
	author = {Sindhwani, Shyamli and Lutteroth, Christof and Weber, Gerald},
	month = may,
	year = {2019},
	pages = {1--13}
}

@inproceedings{turner_gazerst_2015,
	address = {Seoul Republic of Korea},
	title = {Gaze+{RST}: {Integrating} {Gaze} and {Multitouch} for {Remote} {Rotate}-{Scale}-{Translate} {Tasks}},
	isbn = {978-1-4503-3145-6},
	shorttitle = {Gaze+{RST}},
	url = {https://dl.acm.org/doi/10.1145/2702123.2702355},
	doi = {10.1145/2702123.2702355},
	language = {en},
	urldate = {2025-09-01},
	booktitle = {Proceedings of the 33rd {Annual} {ACM} {Conference} on {Human} {Factors} in {Computing} {Systems}},
	publisher = {ACM},
	author = {Turner, Jayson and Alexander, Jason and Bulling, Andreas and Gellersen, Hans},
	month = apr,
	year = {2015},
	pages = {4179--4188}
}

@inproceedings{sidenmark_bimodalgaze_2020,
	address = {Stuttgart Germany},
	title = {{BimodalGaze}: {Seamlessly} {Refined} {Pointing} with {Gaze} and {Filtered} {Gestural} {Head} {Movement}},
	isbn = {978-1-4503-7133-9},
	shorttitle = {{BimodalGaze}},
	url = {https://dl.acm.org/doi/10.1145/3379155.3391312},
	doi = {10.1145/3379155.3391312},
	language = {en},
	urldate = {2025-09-01},
	booktitle = {{ACM} {Symposium} on {Eye} {Tracking} {Research} and {Applications}},
	publisher = {ACM},
	author = {Sidenmark, Ludwig and Mardanbegi, Diako and Gomez, Argenis Ramirez and Clarke, Christopher and Gellersen, Hans},
	month = jun,
	year = {2020},
	pages = {1--9}
}

@inproceedings{stellmach_still_2013,
	address = {Paris France},
	title = {Still looking: investigating seamless gaze-supported selection, positioning, and manipulation of distant targets},
	isbn = {978-1-4503-1899-0},
	shorttitle = {Still looking},
	url = {https://dl.acm.org/doi/10.1145/2470654.2470695},
	doi = {10.1145/2470654.2470695},
	language = {en},
	urldate = {2025-09-01},
	booktitle = {Proceedings of the {SIGCHI} {Conference} on {Human} {Factors} in {Computing} {Systems}},
	publisher = {ACM},
	author = {Stellmach, Sophie and Dachselt, Raimund},
	month = apr,
	year = {2013},
	pages = {285--294}
}

@inproceedings{kyto_pinpointing_2018,
	address = {Montreal QC Canada},
	title = {Pinpointing: {Precise} {Head}- and {Eye}-{Based} {Target} {Selection} for {Augmented} {Reality}},
	isbn = {978-1-4503-5620-6},
	shorttitle = {Pinpointing},
	url = {https://dl.acm.org/doi/10.1145/3173574.3173655},
	doi = {10.1145/3173574.3173655},
	language = {en},
	urldate = {2025-09-01},
	booktitle = {Proceedings of the 2018 {CHI} {Conference} on {Human} {Factors} in {Computing} {Systems}},
	publisher = {ACM},
	author = {Kytö, Mikko and Ens, Barrett and Piumsomboon, Thammathip and Lee, Gun A. and Billinghurst, Mark},
	month = apr,
	year = {2018},
	pages = {1--14}
}

@inproceedings{cai_gazeswipe_2025,
	address = {Yokohama Japan},
	title = {{GazeSwipe}: {Enhancing} {Mobile} {Touchscreen} {Reachability} through {Seamless} {Gaze} and {Finger}-{Swipe} {Integration}},
	isbn = {979-8-4007-1394-1},
	shorttitle = {{GazeSwipe}},
	url = {https://dl.acm.org/doi/10.1145/3706598.3713739},
	doi = {10.1145/3706598.3713739},
	language = {en},
	urldate = {2025-09-01},
	booktitle = {Proceedings of the 2025 {CHI} {Conference} on {Human} {Factors} in {Computing} {Systems}},
	publisher = {ACM},
	author = {Cai, Zhuojiang and Hong, Jingkai and Wang, Zhimin and Lu, Feng},
	month = apr,
	year = {2025},
	pages = {1--14}
}

@inproceedings{stellmach_look_2012,
	address = {Austin Texas USA},
	title = {Look \& touch: gaze-supported target acquisition},
	isbn = {978-1-4503-1015-4},
	shorttitle = {Look \& touch},
	url = {https://dl.acm.org/doi/10.1145/2207676.2208709},
	doi = {10.1145/2207676.2208709},
	language = {en},
	urldate = {2025-09-01},
	booktitle = {Proceedings of the {SIGCHI} {Conference} on {Human} {Factors} in {Computing} {Systems}},
	publisher = {ACM},
	author = {Stellmach, Sophie and Dachselt, Raimund},
	month = may,
	year = {2012},
	pages = {2981--2990}
}

@inproceedings{pfeuffer_gaze_2016,
	address = {Tokyo Japan},
	title = {Gaze and {Touch} {Interaction} on {Tablets}},
	isbn = {978-1-4503-4189-9},
	url = {https://dl.acm.org/doi/10.1145/2984511.2984514},
	doi = {10.1145/2984511.2984514},
	language = {en},
	urldate = {2025-09-01},
	booktitle = {Proceedings of the 29th {Annual} {Symposium} on {User} {Interface} {Software} and {Technology}},
	publisher = {ACM},
	author = {Pfeuffer, Ken and Gellersen, Hans},
	month = oct,
	year = {2016},
	pages = {301--311}
}

@inproceedings{fares_can_2013,
	address = {Paris France},
	title = {Can we beat the mouse with {MAGIC}?},
	isbn = {978-1-4503-1899-0},
	url = {https://dl.acm.org/doi/10.1145/2470654.2466183},
	doi = {10.1145/2470654.2466183},
	language = {en},
	urldate = {2025-09-01},
	booktitle = {Proceedings of the {SIGCHI} {Conference} on {Human} {Factors} in {Computing} {Systems}},
	publisher = {ACM},
	author = {Fares, Ribel and Fang, Shaomin and Komogortsev, Oleg},
	month = apr,
	year = {2013},
	pages = {1387--1390}
}

@inproceedings{jacob_what_1990,
	address = {Seattle, Washington, United States},
	title = {What you look at is what you get: eye movement-based interaction techniques},
	copyright = {https://www.acm.org/publications/policies/copyright\_policy\#Background},
	isbn = {978-0-201-50932-8},
	shorttitle = {What you look at is what you get},
	url = {http://portal.acm.org/citation.cfm?doid=97243.97246},
	doi = {10.1145/97243.97246},
	language = {en},
	urldate = {2025-09-01},
	booktitle = {Proceedings of the {SIGCHI} conference on {Human} factors in computing systems {Empowering} people - {CHI} '90},
	publisher = {ACM Press},
	author = {Jacob, Robert J. K.},
	year = {1990},
	pages = {11--18}
}

@article{jeong_gazehand_2023,
	title = {{GazeHand}: {A} {Gaze}-{Driven} {Virtual} {Hand} {Interface}},
	volume = {11},
	issn = {2169-3536},
	shorttitle = {{GazeHand}},
	url = {https://ieeexplore.ieee.org/document/10332195/},
	doi = {10.1109/ACCESS.2023.3337372},
	urldate = {2025-09-01},
	journal = {IEEE Access},
	author = {Jeong, Jaejoon and Kim, Soo-Hyung and Yang, Hyung-Jeong and Lee, Gun A. and Kim, Seungwon},
	year = {2023},
	pages = {133703--133716}
}

@inproceedings{turner_eye_2013,
	address = {Berlin, Heidelberg},
	title = {Eye {Pull}, {Eye} {Push}: {Moving} {Objects} between {Large} {Screens} and {Personal} {Devices} with {Gaze} and {Touch}},
	isbn = {978-3-642-40480-1},
	shorttitle = {Eye {Pull}, {Eye} {Push}},
	doi = {10.1007/978-3-642-40480-1_11},
	language = {en},
	booktitle = {Human-{Computer} {Interaction} – {INTERACT} 2013},
	publisher = {Springer},
	author = {Turner, Jayson and Alexander, Jason and Bulling, Andreas and Schmidt, Dominik and Gellersen, Hans},
	editor = {Kotzé, Paula and Marsden, Gary and Lindgaard, Gitte and Wesson, Janet and Winckler, Marco},
	year = {2013},
	pages = {170--186}
}

@inproceedings{velloso_empirical_2015,
	address = {Cham},
	title = {An {Empirical} {Investigation} of {Gaze} {Selection} in {Mid}-{Air} {Gestural} {3D} {Manipulation}},
	isbn = {978-3-319-22668-2},
	doi = {10.1007/978-3-319-22668-2_25},
	language = {en},
	booktitle = {Human-{Computer} {Interaction} – {INTERACT} 2015},
	publisher = {Springer International Publishing},
	author = {Velloso, Eduardo and Turner, Jayson and Alexander, Jason and Bulling, Andreas and Gellersen, Hans},
	editor = {Abascal, Julio and Barbosa, Simone and Fetter, Mirko and Gross, Tom and Palanque, Philippe and Winckler, Marco},
	year = {2015},
	pages = {315--330}
}

@article{shi_exploring_2023,
	title = {Exploring {Gaze}-assisted and {Hand}-based {Region} {Selection} in {Augmented} {Reality}},
	volume = {7},
	issn = {2573-0142},
	url = {https://dl.acm.org/doi/10.1145/3591129},
	doi = {10.1145/3591129},
	language = {en},
	number = {ETRA},
	urldate = {2025-09-01},
	journal = {Proc. ACM Hum.-Comput. Interact.},
	author = {Shi, Rongkai and Wei, Yushi and Qin, Xueying and Hui, Pan and Liang, Hai-Ning},
	month = may,
	year = {2023},
	pages = {1--19}
}

@inproceedings{tanriverdi_interacting_2000,
	address = {The Hague The Netherlands},
	title = {Interacting with eye movements in virtual environments},
	isbn = {978-1-58113-216-8},
	url = {https://dl.acm.org/doi/10.1145/332040.332443},
	doi = {10.1145/332040.332443},
	language = {en},
	urldate = {2025-09-01},
	booktitle = {Proceedings of the {SIGCHI} conference on {Human} {Factors} in {Computing} {Systems}},
	publisher = {ACM},
	author = {Tanriverdi, Vildan and Jacob, Robert J. K.},
	month = apr,
	year = {2000},
	pages = {265--272}
}

@inproceedings{turner_eye_2013-1,
	address = {Luleå Sweden},
	title = {Eye drop: an interaction concept for gaze-supported point-to-point content transfer},
	isbn = {978-1-4503-2648-3},
	shorttitle = {Eye drop},
	url = {https://dl.acm.org/doi/10.1145/2541831.2541868},
	doi = {10.1145/2541831.2541868},
	language = {en},
	urldate = {2025-09-01},
	booktitle = {Proceedings of the 12th {International} {Conference} on {Mobile} and {Ubiquitous} {Multimedia}},
	publisher = {ACM},
	author = {Turner, Jayson and Bulling, Andreas and Alexander, Jason and Gellersen, Hans},
	month = dec,
	year = {2013},
	pages = {1--4}
}

@inproceedings{turner_cross-device_2014,
	address = {Safety Harbor Florida},
	title = {Cross-device gaze-supported point-to-point content transfer},
	isbn = {978-1-4503-2751-0},
	url = {https://dl.acm.org/doi/10.1145/2578153.2578155},
	doi = {10.1145/2578153.2578155},
	language = {en},
	urldate = {2025-09-01},
	booktitle = {Proceedings of the {Symposium} on {Eye} {Tracking} {Research} and {Applications}},
	publisher = {ACM},
	author = {Turner, Jayson and Bulling, Andreas and Alexander, Jason and Gellersen, Hans},
	month = mar,
	year = {2014},
	pages = {19--26}
}

@inproceedings{zhai_manual_1999,
	address = {Pittsburgh, Pennsylvania, United States},
	title = {Manual and gaze input cascaded ({MAGIC}) pointing},
	copyright = {https://www.acm.org/publications/policies/copyright\_policy\#Background},
	isbn = {978-0-201-48559-2},
	url = {http://portal.acm.org/citation.cfm?doid=302979.303053},
	doi = {10.1145/302979.303053},
	language = {en},
	urldate = {2025-09-01},
	booktitle = {Proceedings of the {SIGCHI} conference on {Human} factors in computing systems the {CHI} is the limit - {CHI} '99},
	publisher = {ACM Press},
	author = {Zhai, Shumin and Morimoto, Carlos and Ihde, Steven},
	year = {1999},
	pages = {246--253}
}

@inproceedings{pfeuffer_gaze-touch_2014,
	address = {Honolulu Hawaii USA},
	title = {Gaze-touch: combining gaze with multi-touch for interaction on the same surface},
	isbn = {978-1-4503-3069-5},
	shorttitle = {Gaze-touch},
	url = {https://dl.acm.org/doi/10.1145/2642918.2647397},
	doi = {10.1145/2642918.2647397},
	language = {en},
	urldate = {2025-09-01},
	booktitle = {Proceedings of the 27th annual {ACM} symposium on {User} interface software and technology},
	publisher = {ACM},
	author = {Pfeuffer, Ken and Alexander, Jason and Chong, Ming Ki and Gellersen, Hans},
	month = oct,
	year = {2014},
	pages = {509--518}
}

@inproceedings{pfeuffer_gaze_2017,
	address = {Brighton United Kingdom},
	title = {Gaze + pinch interaction in virtual reality},
	isbn = {978-1-4503-5486-8},
	url = {https://dl.acm.org/doi/10.1145/3131277.3132180},
	doi = {10.1145/3131277.3132180},
	language = {en},
	urldate = {2025-09-01},
	booktitle = {Proceedings of the 5th {Symposium} on {Spatial} {User} {Interaction}},
	publisher = {ACM},
	author = {Pfeuffer, Ken and Mayer, Benedikt and Mardanbegi, Diako and Gellersen, Hans},
	month = oct,
	year = {2017},
	pages = {99--108}
}

@article{li_evaluating_2024,
	title = {Evaluating the effects of user motion and viewing mode on target selection in augmented reality},
	volume = {191},
	issn = {10715819},
	url = {https://linkinghub.elsevier.com/retrieve/pii/S1071581924001113},
	doi = {10.1016/j.ijhcs.2024.103327},
	language = {en},
	urldate = {2025-09-01},
	journal = {International Journal of Human-Computer Studies},
	author = {Li, Yang and Liu, Juan and Huang, Jin and Zhang, Yang and Peng, Xiaolan and Bian, Yulong and Tian, Feng},
	month = nov,
	year = {2024},
	pages = {103327}
}

@article{SuEnhanced2025,
author = {Mengyuan Sun and Dongliang Guo and Fengyi Yang and Yapeng Liu},
title = {Enhanced Target Selection in Dense VR Environments Using Single-Hand Frozen Gestures},
journal = {International Journal of Human–Computer Interaction},
volume = {0},
number = {0},
pages = {1--10},
year = {2025},
publisher = {Taylor \& Francis},
doi = {10.1080/10447318.2025.2474473},
URL = { 
        https://doi.org/10.1080/10447318.2025.2474473
    }
}

@article{xu_eyeexpand,
	title = {{EyeExpand}: {A} {Low}-{Burden} and {Accurate} {3D} {Object} {Selection} {Method} {With} {Gaze} and {Raycasting}},
	volume = {n/a},
	issn = {1467-8659},
	shorttitle = {{EyeExpand}},
	url = {https://onlinelibrary.wiley.com/doi/abs/10.1111/cgf.70144},
	doi = {10.1111/cgf.70144},
	language = {en},
	year = {2025},
	journal = {Computer Graphics Forum},
	author = {Xu, X. and He, Y. and Ge, Y. and Zheng, Z.},
	note = {\_eprint: https://onlinelibrary.wiley.com/doi/pdf/10.1111/cgf.70144},
	pages = {e70144}
}

@article{shin_using_2025,
	title = {Using {Augmented} {Reality} on the go: {Understanding} the effects of mobility on user performance and subjective workload across eye, head, and hand ray pointing},
	volume = {204},
	issn = {10715819},
	shorttitle = {Using {Augmented} {Reality} on the go},
	url = {https://linkinghub.elsevier.com/retrieve/pii/S1071581925001545},
	doi = {10.1016/j.ijhcs.2025.103597},
	language = {en},
	urldate = {2025-09-01},
	journal = {International Journal of Human-Computer Studies},
	author = {Shin, Yonghwan and Esteves, Augusto and Oakley, Ian},
	month = oct,
	year = {2025},
	pages = {103597}
}

@inproceedings{kumar_eyepoint_2007,
	address = {San Jose California USA},
	title = {{EyePoint}: practical pointing and selection using gaze and keyboard},
	isbn = {978-1-59593-593-9},
	shorttitle = {{EyePoint}},
	url = {https://dl.acm.org/doi/10.1145/1240624.1240692},
	doi = {10.1145/1240624.1240692},
	language = {en},
	urldate = {2025-09-01},
	booktitle = {Proceedings of the {SIGCHI} {Conference} on {Human} {Factors} in {Computing} {Systems}},
	publisher = {ACM},
	author = {Kumar, Manu and Paepcke, Andreas and Winograd, Terry},
	month = apr,
	year = {2007},
	pages = {421--430}
}

@article{niu_smooth_2023,
	title = {Smooth {Pursuit} {Study} on an {Eye}-{Control} {System} for {Continuous} {Variable} {Adjustment} {Tasks}},
	volume = {39},
	issn = {1044-7318},
	url = {https://doi.org/10.1080/10447318.2021.2012979},
	doi = {10.1080/10447318.2021.2012979},
	number = {1},
	urldate = {2025-09-01},
	journal = {International Journal of Human–Computer Interaction},
	author = {Niu, Yafeng and Li, Xin and Yang, Wenjun and Xue, Chengqi and Peng, Ningyue and Jin, Tao},
	month = jan,
	year = {2023},
	note = {Publisher: Taylor \& Francis
\_eprint: https://doi.org/10.1080/10447318.2021.2012979},
	pages = {23--33}
}

@inproceedings{eyelid3d_2022,
author = {Yi, Xin and Qiu, Leping and Tang, Wenjing and Fan, Yehan and Li, Hewu and Shi, Yuanchun},
title = {DEEP: 3D Gaze Pointing in Virtual Reality Leveraging Eyelid Movement},
year = {2022},
isbn = {9781450393201},
publisher = {Association for Computing Machinery},
address = {New York, NY, USA},
url = {https://doi.org/10.1145/3526113.3545673},
doi = {10.1145/3526113.3545673},
booktitle = {Proceedings of the 35th Annual ACM Symposium on User Interface Software and Technology},
articleno = {3},
numpages = {14},
location = {Bend, OR, USA},
series = {UIST '22}
}

@inproceedings{GonzalezXDTK2024,
    author    = {Gonzalez, Eric J. and Patel, Khushman and Ahuja, Karan and Gonzalez-Franco, Mar},
    title     = {XDTK: A Cross-Device Toolkit for Input \& Interaction in XR},
    booktitle = {2024 IEEE Conference on Virtual Reality and 3D User Interfaces Abstracts and Workshops (VRW)},
    year      = {2024},
    address   = {Orlando, Florida},
    url       = {https://github.com/google/xdtk}
}

@ARTICLE{Weightedpointer,
  author={Sidenmark, Ludwig and Parent, Mark and Wu, Chi-Hao and Chan, Joannes and Glueck, Michael and Wigdor, Daniel and Grossman, Tovi and Giordano, Marcello},
  journal={IEEE Transactions on Visualization and Computer Graphics}, 
  title={Weighted Pointer: Error-aware Gaze-based Interaction through Fallback Modalities}, 
  year={2022},
  volume={28},
  number={11},
  pages={3585-3595},
  doi={10.1109/TVCG.2022.3203096}}

@inproceedings{lookandlean,
author = {\v{S}pakov, Oleg and Isokoski, Poika and Majaranta, P\"{a}ivi},
title = {Look and lean: accurate head-assisted eye pointing},
year = {2014},
isbn = {9781450327510},
publisher = {Association for Computing Machinery},
address = {New York, NY, USA},
url = {https://doi.org/10.1145/2578153.2578157},
doi = {10.1145/2578153.2578157},
booktitle = {Proceedings of the Symposium on Eye Tracking Research and Applications},
pages = {35–42},
numpages = {8},
location = {Safety Harbor, Florida},
series = {ETRA '14}
}

@INPROCEEDINGS{H2H,
  author={Patel, Khushman and Phadnis, Vrushank and Gonzalez, Eric J and Gellersen, Hans and Pfeuffer, Ken and Gonzalez-Franco, Mar},
  booktitle={2025 IEEE Conference on Virtual Reality and 3D User Interfaces Abstracts and Workshops (VRW)}, 
  title={H2E: Hand, Head, Eye a Multimodal Cascade of Natural Inputs}, 
  year={2025},
  volume={},
  number={},
  pages={84-89},
  doi={10.1109/VRW66409.2025.00026}
}

@inproceedings{gazebutton,
author = {Rivu, Sheikh and Abdrabou, Yasmeen and Mayer, Thomas and Pfeuffer, Ken and Alt, Florian},
title = {GazeButton: enhancing buttons with eye gaze interactions},
year = {2019},
isbn = {9781450367097},
publisher = {Association for Computing Machinery},
address = {New York, NY, USA},
url = {https://doi.org/10.1145/3317956.3318154},
doi = {10.1145/3317956.3318154},
booktitle = {Proceedings of the 11th ACM Symposium on Eye Tracking Research \& Applications},
articleno = {73},
numpages = {7},
location = {Denver, Colorado},
series = {ETRA '19}
}

@inproceedings{jang2017modeling,
  title={Modeling cumulative arm fatigue in mid-air interaction based on perceived exertion and kinetics of arm motion},
  author={Jang, Sujin and Stuerzlinger, Wolfgang and Ambike, Satyajit and Ramani, Karthik},
  booktitle={Proceedings of the 2017 CHI conference on human factors in computing systems},
  pages={3328--3339},
  year={2017}
}

@inproceedings{sonohaptic,
author = {Cho, Hyunsung and Sendhilnathan, Naveen and Nebeling, Michael and Wang, Tianyi and Padmanabhan, Purnima and Browder, Jonathan and Lindlbauer, David and Jonker, Tanya R. and Todi, Kashyap},
title = {SonoHaptics: An Audio-Haptic Cursor for Gaze-Based Object Selection in XR},
year = {2024},
isbn = {9798400706288},
publisher = {Association for Computing Machinery},
address = {New York, NY, USA},
url = {https://doi.org/10.1145/3654777.3676384},
doi = {10.1145/3654777.3676384},
booktitle = {Proceedings of the 37th Annual ACM Symposium on User Interface Software and Technology},
articleno = {125},
numpages = {19},
location = {Pittsburgh, PA, USA},
series = {UIST '24}
}

@inproceedings{haptic_eyestyping,
author = {Gupta, Aakar and Sendhilnathan, Naveen and Hartcher-O'Brien, Jess and Pezent, Evan and Benko, Hrvoje and Jonker, Tanya R.},
title = {Investigating Eyes-away Mid-air Typing in Virtual Reality using Squeeze haptics-based Postural Reinforcement},
year = {2023},
isbn = {9781450394215},
publisher = {Association for Computing Machinery},
address = {New York, NY, USA},
url = {https://doi.org/10.1145/3544548.3581467},
doi = {10.1145/3544548.3581467},
booktitle = {Proceedings of the 2023 CHI Conference on Human Factors in Computing Systems},
articleno = {230},
numpages = {11},
location = {Hamburg, Germany},
series = {CHI '23}
}

@incollection{nasatlx,
  title        = {Development of NASA-TLX (Task Load Index): Results of Empirical and Theoretical Research},
  author       = {Hart, Sandra G. and Staveland, Lowell E.},
  booktitle    = {Advances in Psychology},
  editor       = {Hancock, Peter A. and Meshkati, Najmedin},
  volume       = {52},
  pages        = {139--183},
  year         = {1988},
  publisher    = {North-Holland},
  isbn         = {9780444703880},
  issn         = {0166-4115},
  doi          = {10.1016/S0166-4115(08)62386-9},
  url          = {https://www.sciencedirect.com/science/article/pii/S0166411508623869}
}

@inproceedings{gazeonthego,
author = {Manakhov, Pavel and Sidenmark, Ludwig and Pfeuffer, Ken and Gellersen, Hans},
title = {Gaze on the Go: Effect of Spatial Reference Frame on Visual Target Acquisition During Physical Locomotion in Extended Reality},
year = {2024},
isbn = {9798400703300},
publisher = {Association for Computing Machinery},
address = {New York, NY, USA},
url = {https://doi.org/10.1145/3613904.3642915},
doi = {10.1145/3613904.3642915},
booktitle = {Proceedings of the 2024 CHI Conference on Human Factors in Computing Systems},
articleno = {373},
numpages = {16},
location = {Honolulu, HI, USA},
series = {CHI '24}
}

@techreport{ISO9241-9,
  author       = {{International Organization for Standardization}},
  title        = {ISO 9241-9: Ergonomic requirements for office work with visual display terminals (VDTs) -- Part 9: Requirements for non-keyboard input devices},
  institution  = {International Organization for Standardization},
  number       = {ISO 9241-9},
  year         = {2000}
}

@inproceedings{mobility_smartwatch,
author = {Dobbelstein, David and Haas, Gabriel and Rukzio, Enrico},
title = {The effects of mobility, encumbrance, and (non-)dominant hand on interaction with smartwatches},
year = {2017},
isbn = {9781450351881},
publisher = {Association for Computing Machinery},
address = {New York, NY, USA},
url = {https://doi.org/10.1145/3123021.3123033},
doi = {10.1145/3123021.3123033},
booktitle = {Proceedings of the 2017 ACM International Symposium on Wearable Computers},
pages = {90–93},
numpages = {4},
location = {Maui, Hawaii},
series = {ISWC '17}
}

@article{headar,
author = {Yang, Xiaoying and Wang, Xue and Dong, Gaofeng and Yan, Zihan and Srivastava, Mani and Hayashi, Eiji and Zhang, Yang},
title = {Headar: Sensing Head Gestures for Confirmation Dialogs on Smartwatches with Wearable Millimeter-Wave Radar},
year = {2023},
issue_date = {September 2023},
publisher = {Association for Computing Machinery},
address = {New York, NY, USA},
volume = {7},
number = {3},
url = {https://doi.org/10.1145/3610900},
doi = {10.1145/3610900},
journal = {Proc. ACM Interact. Mob. Wearable Ubiquitous Technol.},
month = sep,
articleno = {138},
numpages = {28}
}

\appendix
\section{User Feedback}
\label{a:userfeedback}
\paragraph{\textit{GazeTap}}Participants generally described \textit{GazeTap} as the easiest and fastest technique that required minimal effort. However, many reported frustration when aligning the cursor precisely on targets and expressed lower confidence in the final placement. To compensate, users often resorted to subtle head movements or tensing their eyes for fine alignment. For example, P18 noted, \textit{“Staring at a single target made my eyes tired,”} and P20 added, \textit{“I had to stare and try not to blink, so my eyes got tired.”} Several participants also remarked that while gaze quickly brought the cursor in the right direction, achieving precision was difficult (e.g., P9: \textit{“I often moved my head to line things up.”}).

\paragraph{\textit{GazePinch}}In a \textit{Stationary}, \textit{GazePinch} was perceived as natural and direct, even for participants without prior AR/VR experience, making it feel intuitive overall. Participants highlighted arm fatigue for \textit{GazePinch}: (P5: \textit{“While it felt straightforward and ‘pretty easy’ when seated, because the simple reach-out made initial acquisition quick. Keeping my arm extended was tiring, and I disliked using it while \textit{Motion-Induced}.”}) Some also described the repeated reach-out/re-attempt cycles as mentally taxing when they had to correct placement. Most participants found it effortful and reported getting tired more quickly in \textit{Motion-Induced}. Several noted that when strict accuracy was not required, they preferred pinch for its straightforwardness (P7: \textit{“I preferred pinch when precision wasn’t critical”}; P16 echoed this preference).

\paragraph{GazeTune} \textit{GazeTune} was initially less intuitive and took longer than \textit{GazeTap} or \textit{GazePinch}, but most participants reported that it became easy to use after a few trials and was valued for its accuracy. Several emphasized lower attentional demand: rather than holding a rigid gaze on the target, they could refine with a finger while maintaining a softer focus (P3: \textit{“accurate, and I don’t need to focus on the target”}). Participants also appreciated reduced head movement and the ability to make comfortable finger adjustments even while \textit{Motion-Induced} (P2: \textit{“I could keep my head steady and move my finger comfortably”}). And some participants framed it as a time accuracy trade-off (P6: \textit{“It costs time, but I’m willing to accept that”}. P11: \textit{“The refinement gives high accuracy, but it costs time”}). Once familiar, some found it to be reliably precise, but it took a slightly longer time (P9), while others felt it worked better than pure eye gaze when gaze errors occurred (P20).
\end{document}